\documentclass[a4paper,11pt]{article}
\usepackage{jheppub}
\usepackage{mathrsfs}
\usepackage{amsfonts}
\usepackage{setspace}
\usepackage{cellspace}
\usepackage{amsmath,amssymb,bm}
\usepackage[colorlinks=true,linkcolor=blue]{hyperref}
\usepackage{xcolor}
\usepackage{epsfig}
\usepackage{slashed}
\usepackage{caption}
\usepackage{hhline,multirow,tabularx}  % for nicer tables
\usepackage{dcolumn}    % align table columns on decimal point
\usepackage{url}        % for URL addresses
\usepackage{braket} 
\definecolor{cyan}{rgb}{0.0, 1.0, 1.0}
\definecolor{applegreen}{rgb}{0.55, 0.71, 0.0}
\definecolor{arylideyellow}{rgb}{0.91, 0.84, 0.42}
\definecolor{bananayellow}{rgb}{1.0, 0.88, 0.21}
\definecolor{burlywood}{rgb}{0.87, 0.72, 0.53}
\definecolor{buff}{rgb}{0.94, 0.86, 0.51}
\definecolor{blond}{rgb}{0.98, 0.94, 0.75}
\definecolor{bisque}{rgb}{1.0, 0.89, 0.77}
\definecolor{bananamania}{rgb}{0.98, 0.91, 0.71}
\definecolor{apricot}{rgb}{0.98, 0.81, 0.69}
\definecolor{almond}{rgb}{0.94, 0.87, 0.8}

\newenvironment{bbox}
{\par\smallskip\centering\begin{lrbox}{0}%
\begin{minipage}[c]{0.99\textwidth}}
{\end{minipage}\end{lrbox}%
\framebox[0.99\textwidth]{\usebox{0}}%
\par\medskip}

\title{Renormalon cancellation and linear power correction to threshold-like asymptotics of space-like parton correlators}
\author[a]{Yizhuang Liu}
\author[b]{Yushan Su}

\affiliation[a]{Institute of Theoretical Physics,
Jagiellonian University, 30-348 Kraków, Poland}
\affiliation[b ]{Department of Physics, University of Maryland, College Park, MD 20742}

\emailAdd{yizhuang.liu@uj.edu.pl}
\emailAdd{ysu12345@umd.edu}

\abstract {In this paper, we show that the common hard kernel of double-log-type or threshold-type factorization for certain space-like parton correlators that arise in the context of lattice parton distributions, the {\it heavy-light Sudakov hard kernel} 
, has linear infrared (IR) renormalon. We explicitly demonstrate how this IR renormalon correlates with ultraviolet (UV) renormalons of next-to-leading power operators in two explicit examples: threshold asymptotics of space-like  quark-bilinear coefficient functions and transverse momentum dependent (TMD) factorization of quasi wave function amplitude. Theoretically, the pattern of renormalon cancellation complies with general expectations to marginal asymptotics in the UV limit.  Practically, this linear renormalon explains the slow convergence of imaginary parts observed in lattice extraction of the Collins-Soper kernel and signals the relevance of next-to-leading power contributions. Fully factorized, fully controlled threshold asymptotic expansion for space-like  quark-bilinear coefficient functions in coordinate and moment space has also been proposed.} 
\date{\today}

\begin{document}
\maketitle
\flushbottom

\section{Introduction}
It is known that perturbative series induced by marginal perturbations to UV or IR conformal field theories (CFTs) are tricky and subtle. On one hand, fixed-order Feynman-diagram-based calculations suffer from uncontrolled large logarithms that must be resummed using renormalization group equation (RGE) or whatever structures to become controlled asymptotic expansions. On the other hand, even after resummation, normally the perturbative series can only be a divergent, Borel non-summable asymptotic series. Physically, this reflects the fact that in the presence of marginal perturbation, the decoupling between UV and IR scales is very slow, and the Borel non-summability is the price one pays for decoupling asymptotically. 

In fact, Borel non-summable logarithmic asymptotic series is not unusual in nature. A famous example from mathematics is the prime number theorem (PNT), stating that the asymptotic distribution $\pi(x)$ of prime numbers takes the form 
\begin{align}\label{eq:PNT}
\pi (x)\sim \frac{x}{\ln x}\sum_{n=0}^{\infty} \frac{n!}{\ln^n x} \ ,
\end{align}
in the $x\rightarrow \infty$ limit, which is similar in shape to a perturbative series with ``one-loop running coupling constant'' $\alpha(x)=\frac{1}{\ln x}$. One can check that the Borel resummation of the above series has a pole at $t=1$, and different prescriptions, such as the principle value or $\pm i0$ prescriptions lead to $O(1)$ ambiguity
\begin{align}
{\rm PV}\int_{0}^{x}\frac{dt}{\ln t}-\int_{0}^{x} \frac{dt}{\ln t \pm i0}= \pm \pi i \ .
\end{align}
Can one read from this ambiguity any useful information regarding ``power corrections'' to the PNT? The answer is unfortunately no.  In fact, one knows that the true ``next-to-leading-power'' correction comes from non-trivial zeros of the Riemann zeta function and is of the order $\sqrt{x}$ (assuming Riemann hypothesis)
\begin{align}
\pi(x)-{\rm PV}\int_{0}^{x}\frac{dt}{\ln t}={\cal O}(\sqrt{x}) \ .
\end{align}
In such a case, the Borel singularity at ``leading power'' carries no useful information regarding the power corrections at all. 

Given this example, one may ask the following question: near UV fixed point of local quantum field theories (QFTs), is it also true that marginal perturbative series for IR safe quantities, even with their singularity structures in the Borel plane completely decoded, still can not tell where the non-trivial power corrections are?  The answer to this question is fortunately {\it no}. In fact, although the perturbative series in QFT is usually much more complicated than Eq.~(\ref{eq:PNT}) and in prior one can not even see where the Borel singularity is, due to the special context of local QFTs, important understandings can still be formed. In particular, the {\it perturbative renomalon}~\cite{tHooft:1977xjm,Parisi:1978az,Parisi:1978bj} is widely believed to be a crucial structure that can provide useful information regarding the high-order behavior of perturbative series as well as power corrections. 

Based on our understanding,  positive arguments and evidences in favor of perturbative renormalons can be summarized as below. 
\begin{enumerate}
    \item First of all, they are caused by a small amount of diagrams (such as the {\it bubble chain diagram} for QCD) which allow rigorous analysis as well as explicit calculations to show that they indeed lead to non-alternating factorial growths of the form $(\frac{\beta_0}{2 k})^n n!$ which would become Borel singularity at $t=\frac{2 k}{\beta_0}$ if not miraculously canceled by other diagrams. 
    \item Second, at the level of Feynman integrals, the factorial growth is caused by running inside loop integrals which exactly drives the loop momenta towards the ``dangerous direction'', such as the small momentum region in QCD,  suggesting natural correlation with power corrections or non-perturbative contributions.
    \item  Third, in several examples where the high power contributions are strongly believed to be caused by high-dimensional operators in the operator product expansion (OPE), a characteristic structure for CFTs that are very likely to be inherited in some form in their marginal deformations as well, the IR renormalon in the leading power can be shown to correlate~\cite{Shifman:1978bx,David:1983gz,David:1985xj,Mueller:1984vh,Ji:1994md,Beneke:1998ui,Beneke:1998eq,Shifman:2013uka,Dunne:2015eoa} with UV renormalons caused by ambiguities in defining these high dimensional operators.
    \item Fourth, in all the integrable {\it none-supersymmetric} asymptotically safe 2D massive QFTs that have been analyzed using the thermodynamical Bethe's ansatz (TBA) method so far~\cite{Marino:2019eym,Bajnok:2021dri,Marino:2021dzn,Marino:2022ykm,Bajnok:2022xgx,Bajnok:2022rtu,Reis:2022tni,Marino:2023epd}, the leading singularities in the right-half Borel plane (for the perturbative series for ground state energy in the large ``chemical potential'' limit) and the place of NLP contribution all agree\footnote{Notice for the $O(4)$ non-linear sigma model and $SU(N)$ chiral principal field models, contribution at the power predicted by the IR renormalon vanishes in the principal value prescription but exists in $\pm i0$ prescriptions~\cite{Marino:2021dzn,Bajnok:2021dri,Reis:2022tni}. For $\beta^{2}\rightarrow 8\pi^-$ massive sine Gordon, the leading IR renormalon is the only Borel singularity in the right-half plane~\cite{Liu:2023tft} and the perturbative series re-summs to full result in the principal value prescription. } with predictions based on leading perturbative IR renormalon. In particular, for the $O(N)$ non-linear sigma models, both the large-$N$ expansion~\cite{David:1982qv,David:1983gz,Novikov:1984ac,David:1985xj,Beneke:1998eq} at ${\cal O}(\frac{1}{N})$ and the exact TBA method agrees~\cite{DiPietro:2021yxb,Marino:2021dzn,Marino:2021six} and works in favor of perturbative renormalon.
    \item Finally, for the lattice perturbation series for the self-energy of a Wilson-line, the asymptotic behavior extracted from high-order numerical calculations agrees with the renormalon-based prediction~\cite{Bauer:2011ws,Bali:2014fea}. 
    \end{enumerate}
 To summarize, the perturbative renormalon, in particular, the bubble-chain-based renormalon analysis for QCD (sometimes called ``large $\beta_0$ approximation'' in literature) is a valuable method to determine leading Borel singularity of QCD perturbative series as well as to probe, or at least, to provide lower-bounds for high power corrections. 

In this paper, based on this understanding, we investigate the Borel singularity using the bubble-chain diagrams, for a particular hard kernel that appears in several factorization formulas in the large momentum effective theory (LaMET) formulation~\cite{Ji:2013dva,Ji:2014gla,Cichy:2018mum,Ji:2020ect} for lattice parton distributions. This hard kernel first appeared in the factorization of quasi-TMDPDFs~\cite{Ji:2014hxa,Ebert:2019okf,Ji:2019ewn,Vladimirov:2020ofp,Ebert:2020gxr,Ebert:2022fmh}, and then in TMD factorization of quasi light-front wave function (LFWF) amplitudes~\cite{Ji:2021znw} and recently in threshold limit of space-like quark-bilinear coefficient functions~\cite{Ji:2023pba}. It is a {\it double-logarithmic type} or {\it threshold-type} hard kernel in the sense that it satisfies an RGE with a logarithmic term in its anomalous dimension of the form $\Gamma_{\rm cusp}(\alpha)\ln \frac{\zeta_z}{\mu^2}$ controlled by the universal light-like cusp anomalous dimension $\Gamma_{\rm cusp}(\alpha)$~\cite{Korchemsky:1987wg,Henn:2019swt,vonManteuffel:2020vjv} . Moreover, its definition is very simple: it is just an on-shell form factor defined purely in PQCD with UV and IR divergences subtracted {\it multiplicatively} in the $\overline {\rm MS}$ scheme, with an incoming on-shell light-quark with four-momenta $p^2=0$ and an outgoing on-shell heavy Wilson-line in space-like direction $n_z^2=-1$. The Lorentz invariant combination $\zeta_z=4(p\cdot n_z)^2 \ne 0$ serves as the only scale of the form factor. For this reason, it is called the {\it heavy-light Sudakov form factor}~\cite{Ji:2023pba} or {\it heavy-light Sudakov hard kernel} due to its similarity with the standard light-light Sudakov form factor~\cite{Sudakov:1954sw,Collins:1989bt,Moch:2005id,Baikov:2009bg} with on-shell external light-partons. On the other hand, since the gauge link can be interpreted as the gauge transformation that transforms a generic gauge to the {\it axial gauge} , this form factor can also be regarded as the gauge-invariant definition of the {\it quark field renormalization factor} in the axial gauge, for which double-logarithmic asymptotics is known long ago~\cite{Sen:1981sd}.  In all three factorization formulas, this form factor is caused by the Sudakov effect around the quark-link vertices and has recently been determined up to NNLO~\cite{Ji:2023pba,delRio:2023pse}.

Naively looking, this hard kernel should contain only quadratic IR renormalon singularities which is the usual case for hard kernels. On the other hand, the presence of a heavy Wilson-line in its definition resembles quantities in heavy-quark effective theory, for which linear renormalon is common. It is the purpose of this paper to show using explicit calculations that the heavy-light Sudakov form factor contains linear IR renormalon in its {\it phase angle}. Moreover, we show that for both the threshold factorization of space-like quark correlator as well as the TMD factorization of quasi-LFWF amplitude, this linear renormalon cancels with corresponding UV renormalons for operators at next-to-leading power (NLP). As such, these examples can also serve as simple demonstration-of-principle for the general pattern of renormalon cancellation in the context of double-logarithmic-type asymptotic expansions.  The case of quasi-LFWF also provides a simple example of the ``OPE non-convergence'' phenomenon~\cite{Shifman:1994yf,Shifman:1998rb}. We hope that these examples will be not only useful for the lattice extraction of partonic distributions and evolution kernels, but also interesting for their pedagogical value to demonstrate the renormalon cancellation/correlation pattern mentioned before. 

We emphasize that the linear renormalon in the heavy-light Sudakov hard kernel requires  ``on-shell'' space-like gauge-link and affects only the double-logarithmic type factorization of quasi-distributions.  Before power expansions that lead to double-log evolution, the gauge links ``cancel among themselves'' at infinity between ``real'' and virtual diagrams, and the linear renormalons are hidden. After the TMD expansion in $\frac{1}{b_{\perp} p^z}$ or threshold expansion in $\frac{1}{zp^z}$, at leading power the cancellation is no-longer perfect, and the linear renormalon can only be cancelled between leading and higher powers. 

The organization of the paper is as follows.
\begin{enumerate}
    \item In Sec.~\ref{sec:HLbubble}, we calculate explicitly the bubble chain diagram for the hard kernel in the $\overline{\rm MS}$ scheme following the method of~\cite{Palanques-Mestre:1983ogz,Beneke:1994sw,Beneke:1995pq,Scimemi:2016ffw}. We show that the imaginary part suffers linear renormalon ambiguity, while the real part is free from linear renormalon. Furthermore, the strength for the linear renormalon ambiguity happens to be the same as that for the linear divergence of pole mass. As a by-product, we also obtained the corresponding $\overline{\rm MS}$ scheme anomalous dimensions for the bubble chain diagram. 
    \item In Sec.~\ref{sec:threshold}, we investigate the role played by this linear renormalon in the threshold asymptotics of space-like ``quasi''~\cite{Ji:2013dva,Ji:2014gla} or ``pseudo''~\cite{Radyushkin:2017cyf,Radyushkin:2017lvu} quark-bilinear coefficient functions in {\it coordinate space} and {\it moment space}. We first review the factorization formalism,  emphasizing the UV nature of the threshold limit in coordinate space and moment space at fixed $z^2$. A fully-factorized controlled expansion in the $N\rightarrow \infty$ limit has been written down explicitly. 
    
    We then use explicit analytical result of bubble chain diagram~\cite{Braun:2018brg} to perform the threshold expansion both in coordinate and moment space. The threshold factorization formula at leading threshold power is verified at the bubble chain level. The linear renormalon cancels between the heavy-light hard kernel at leading threshold power and the NLP threshold soft factor.
We identify the operator definition for the NLP contribution and demonstrate that its linear renormalon is UV. We show that the ``genuine'' IR renormalons related to $z\Lambda_{\rm QCD}$ power corrections remain the same before and after threshold expansion.  We further investigate the ``resurgence'' of full distribution from threshold asymptotics by noticing the correlation between Borel ambiguity of $e^{-i\lambda}$ terms (corresponding to $\alpha \rightarrow 1$ threshold asymptotics) and branch ambiguity of purely algebraic terms (corresponding to $\alpha \rightarrow 0$ asymptotics).
\item In Sec.~\ref{sec:LFWF}, we perform the same analysis to the TMD factorization of quasi-LFWF amplitude. We show that the bubble chain diagram for the quasi-LWFW amplitude, before the TMD expansion, is free from the linear renormalon. After performing the power expansion, the linear renormalon in the heavy-light Sudakov form factor, playing the role of the hard kernel at leading power, cancels with the UV renormalon of NLP soft contribution. We determine the operator definition of the NLP soft factor and show that its UV renormalon is still correlated with linear UV divergence rather than the linear rapidity divergence. We show that the power expansion in this example fails to sum to the full result due to extra exponentially small contributions and comment on features of TMD expansion in $b_\perp$ and $k_\perp$ space.
\item In Sec.~\ref{sec:coulomb}, we calculate the bubble chain diagram for the quark field wave function renormalization in Coulomb gauge and show that it suffers linear renormalon ambiguity as well.  In Sec.~\ref{sec:conclu}, we discuss the practical implications of this linear renormalon to lattice extraction of Collins-Soper kernel and then conclude.
\end{enumerate}
Useful technical details are collected in the Appendices.

\section{Linear renormalon of the heavy-light Sudakov hard kernel}\label{sec:HLbubble}
In this section, we calculate the bubble chain diagram for the heavy-light Sudakov hard kernel, or the common hard kernel of double-log type factorizations in the context of lattice parton distributions. 

As a reminder, the heavy-light Sudakov hard kernel is a quantity defined purely in PQCD, with UV and IR divergences regulated in DR and subtracted {\it multiplicatively} in the $\overline{\rm MS}$ scheme. It is similar to the standard light-light Sudakov hard kernel, but with one of the external light-parton states replaced by an ``on-shell'' gauge link in $n_z^2=-1$ direction. More precisely, one has 
\begin{align}\label{eq:defHLsuda}
\langle n_z|\bar Q_{n_z}(0)\psi(0)|p\rangle &\equiv \langle \Omega| [{\rm sign}(z)\infty n_z,0]\psi(0)|p\rangle \nonumber\\
&\equiv u(p)H_{\rm HL}\bigg(L_z,{\rm sign}(z),\alpha(\mu)\bigg) \ ,
\end{align}
where $|p\rangle$ is an external collinear quark with the momentum $p=(p^z,0,0,p^z)$ and $\psi(0)$ is a quark field. In this paper, we use the notation $[y,x]$ to denote a Wilson line from $x$ to $y$ along the straight line between them. If $y=\infty$, then $[ \pm \infty v,x]$ means a Wilson-line from $x$ to infinity along the direction $ \pm v$. More precisely, one has
\begin{align}
[{\rm sign}(z)\infty n_z ,0]={\rm P}e^{-i g \int_{0}^{{\rm sign}(z) \infty n_z } d x^{\mu} A_{\mu}(x)} \ .
\end{align}
In Eq.~(\ref{eq:defHLsuda}), $p^z=p_z>0$ should be identified as quark-parton's momenta $xP^z$ in factorization formulas, and we use ${\rm sign}(z)$ to denote the direction of the $n^z$. Due to Lorentz invariance as well as the freedom to rescale $n_z \rightarrow \rho n_z$, the natural scale of the form factor is $\zeta_z=\frac{4|p\cdot n_z|^2}{n_z^2}=4 p_z^2$, which enters the argument of the form factor through the logarithm $L_z \equiv \ln \frac{\zeta_z}{\mu^2}$.  Moreover, the $H_{\rm HL}$ depends on the direction of $n_z$, $\sigma={\rm sign}(z)$ only through the combination $L^{\sigma}=\ln \frac{-2ip^z\sigma}{\mu}$
\begin{align}
H_{\rm HL}\bigg(L_z,{\rm sign}(z),\alpha(\mu)\bigg) \equiv H_{\rm HL}\bigg(\ln \frac{-2i p^z{\rm sign}(z) }{\mu},\alpha(\mu)\bigg) \ .
\end{align}
The hard kernel $H_{\rm HL}$ satisfies the standard form of double-log RGE~\cite{Ji:2019ewn,Ji:2021znw,Ji:2023pba}, \begin{align}\label{eq:RGEHLsuda}
2\mu\frac{d}{d\mu} \ln H_{\rm HL}\bigg(\ln \frac{-2i p^z\sigma }{\mu},\alpha(\mu)\bigg)=2\Gamma_{\rm cusp}(\alpha) \ln \frac{-2i p^z \sigma}{\mu} +\tilde {\gamma}_{H}(\alpha) \ .
\end{align}
In the above RGE, besides the universal light-like cusp anomalous dimension $\Gamma_{\rm cusp}$, there is a single-log anomalous dimension $\tilde \gamma_{H}$, which can be expressed as~\cite{Ji:2019ewn,Ji:2021znw,Ji:2023pba}
\begin{align}\label{eq:heavylighsuda}
\tilde \gamma_{H}=2\gamma_{F}+\gamma_V+2\gamma_{HL}-\gamma_s  \ .
\end{align}
In the above, $\gamma_F$ is the UV anomalous dimension for heavy-light current~\cite{Ji:1991pr,Braun:2020ymy}, $\gamma_V$ is the constant part for the anomalous dimension of the standard light-light Sudakov hard kernel~\cite{Becher:2006mr},  $\gamma_s$ is the soft anomalous dimension, which can be defined through UV anomalous dimension of a light-like Wilson-line cusp~\cite{korchemskaya:1992je}, and $\gamma_{HL}$ is the constant part of the UV anomalous dimension of a heavy-light Wilson-line cusp originally calculated in Ref.~\cite{Korchemsky:1992xv}. They are known to NNNLO. 

Given the RGE, it is possible to introduce the RGE re-summed form of the hard kernel. For this purpose, it is convenient to introduce
\begin{align}\label{eq:hardabs}
 &H\bigg(L_z,\alpha(\mu)\bigg) \equiv \big|H_{\rm HL}\big|^2\bigg(\ln \frac{-2i p^z \sigma }{\mu},\alpha(\mu) \bigg)\ , \\
 & \sigma A\bigg(L_z,\alpha(\mu)\bigg)=2{\rm Arg}\bigg(H_{\rm HL}\left(\ln \frac{-2i p^z \sigma }{\mu},\alpha(\mu)\right)\bigg) \ . \label{eq:defJf}
\end{align}
In terms of the above, the RGE re-summation reads~\cite{Ji:2023pba}
\begin{align}
&H\bigg(L_z,\alpha(\mu)\bigg)=H
\bigg(L_z=0,\alpha(\sqrt{\zeta_z})\bigg)\exp \bigg[2S(\sqrt{\zeta_z},\mu)-a_{H}(\sqrt{\zeta_z},\mu)\bigg] \ , \\ 
&A\bigg(L_z,\alpha(\mu)\bigg)=A\bigg(L_z=0,\alpha(\sqrt{\zeta_z})\bigg)+\pi a_{\Gamma}(\sqrt{\zeta_z},\mu) \ , \label{eq:phaseRG}
\end{align}
where the various RGE resummation factors reads~\cite{Becher:2006mr} 
\begin{align}\label{eq:aGamma}
&S(\nu,\mu)=-\int_{\alpha(\nu)}^{\alpha(\mu)}\frac{\Gamma_{\rm cusp}(\alpha)d\alpha}{\beta(\alpha)}\int_{\alpha(\nu)}^{\alpha}\frac{d\alpha'}{\beta(\alpha')} \ , \\ 
&a_{\rm \Gamma}(\nu,\mu)=-\int_{\alpha(\nu)}^{\alpha(\mu)} d\alpha\frac{\Gamma_{\rm cusp}(\alpha)}{\beta(\alpha)}  \ , \\
&a_{H}(\nu,\mu)=-\int_{\alpha(\nu)}^{\alpha(\mu)} d\alpha\frac{\tilde \gamma_H(\alpha)}{\beta(\alpha)} \ .
\end{align}
Notice that there are imaginary parts of the heavy-light Sudakov form factor, resulting in the $i\pi{\rm sign}(z)$ term in the anomalous dimension and the $a_{\rm \Gamma}$ term in the RGE resummation. As demonstrated in Ref.~\cite{Ji:2021znw}, such an imaginary part can be obtained by applying ``cutting-rules''  $\frac{1}{k^z \pm i0} \rightarrow \mp i\pi\delta(k^z)$ on the eikonal propagators which make one of them on shell. On the other hand, the eikonal propagators are {\it not pinched} at such region. In fact, the $k^z \ll k^0,k_\perp$ region and the $|k^-k^+| \ll k_\perp^2$ region in the loop integrals can be deformed away combining contour deformations $k^- \rightarrow k^- - if(k)$ and $k^+\rightarrow k^+ \pm i f(k)$ with $f(k)>0$. \footnote{For the quasi-TMDPDF, one needs to apply the ``sum over cut'' argument~\cite{Collins:2011zzd} to cancel final state interactions.}. This implies that the heavy-light Sudakov hard kernel as well as the quasi-LFWF amplitudes to be introduced later are deeply inside the natural region of analyticity, in a way similar to the angle-dependent cusp anomalous dimension with a $\frac{i\pi}{2}$ in the hyperbolic angle~\cite{Korchemsky:1987wg}. 
Nevertheless, the presence of an imaginary part or a phase angle in the hard kernel still results in some inconvenience in practical applications~\cite{LatticePartonLPC:2023pdv,Avkhadiev:2023poz}. It is the purpose of this section to show that, the imaginary part of the heavy-light Sudakov hard kernel, or more precisely, the phase angle $A(L_z=0,\alpha(\sqrt{\zeta_z}))$, contains a linear renormalon ambiguity.

\subsection{Bubble-chain diagram for the heavy-light hard kernel}
\begin{figure}[htbp]
    \centering
    \includegraphics[height=5.0cm]{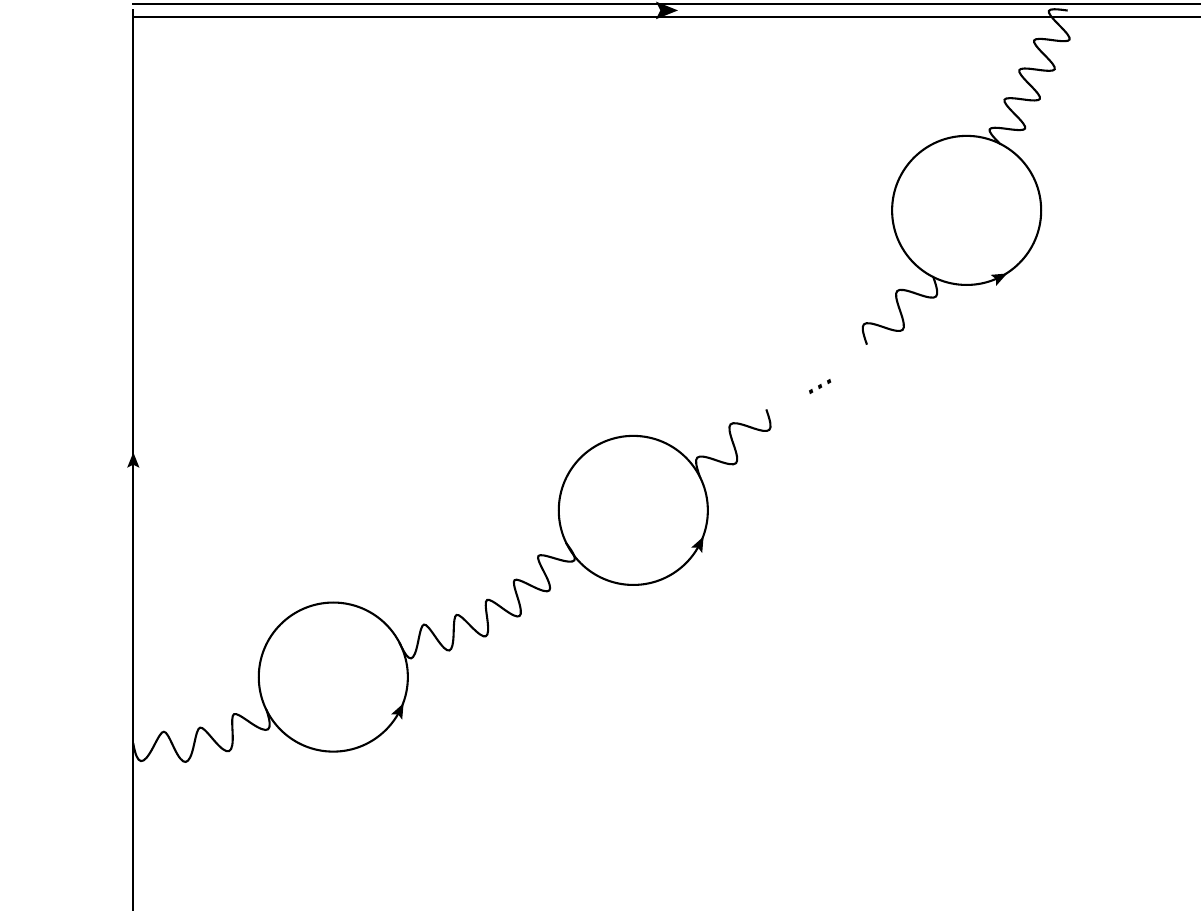}
    \caption{Bubble chain diagram for the heavy light Sudakov hard kernel.}
    \label{fig:hknbubbles}
\end{figure}
Here we calculate the heavy-light bubble chain diagram in Fig.~\ref{fig:hknbubbles}. The momentum space gluon propagator with $n$-bubbles inserted reads
\begin{align}
&\langle A^{\mu}A^{\nu}\rangle=\frac{ig^{\mu\nu}\mu^{2n\epsilon}}{(-k^2-i0)^{1+n\epsilon}}\alpha^n\bigg(\frac{\beta_0}{2}\bigg)^ne^{n\epsilon \gamma_E}\frac{1}{\epsilon^n}f(\epsilon)^n \ , \\
&f(\epsilon)=\frac{6\Gamma^2(2-\epsilon)\Gamma(1+\epsilon)}{\Gamma(4-2\epsilon)}=1+\left(\frac{5}{3}-\gamma_E\right)\epsilon+{\cal O}(\epsilon^2) \ .\label{eq:feps}
\end{align}
Here $\beta_0 = \frac{11 C_A}{6 \pi} - \frac{n_f}{3\pi}$, and $u(p)$ is an on-shell Dirac spinor with $p=(p^z,0,0,p^z)$. Now, the diagram for ${\rm sign}(z)=\sigma=-1$ reads (for $\sigma=1$ one simply fips the imaginary part) 
\begin{align}\label{eq:fnnotation}
V_n u(p)=g^2C_F\alpha^n\bigg(\frac{\beta_0}{2}\bigg)^n\mu_0^{2\epsilon}\mu^{2n\epsilon}e^{n\epsilon \gamma_E}\frac{1}{\epsilon^n}f(\epsilon)^n h_n u(p), \\ 
h_n u(p)=-\int \frac{d^Dk}{(2\pi^D)}\frac{1}{(-k^2-i0)^{1+n\epsilon}}\frac{\slashed{p}+\slashed{k}}{(p+k)^2+i0}\slashed{n_z}\int_{0}^{\infty} d\lambda e^{-ik\cdot n_z \lambda} u(p) \ ,
\end{align}
where $\mu_0^{2\epsilon}=\left(\mu^2\right)^\epsilon\left(\frac{e^{\gamma_E}}{4 \pi}\right)^\epsilon $ and  $n_z=(0,0,0,1)$. After introducing Schwinger parameters and after Wick rotation, one has 
\begin{align}
&h_n=\frac{i}{(4\pi)^{\frac{D}{2}}\Gamma(1+n\epsilon)}\nonumber \\ 
&\times\int_{0}^{\infty}d\lambda \int_{0}^{\infty} d\rho \rho^{-1+(n+1)\epsilon}\int_0^1dx \bar x^{n\epsilon}e^{-\frac{\lambda^2}{4\rho}+i\lambda x p\cdot n_z} \bigg(2(1-x)p\cdot n_z-\frac{i\lambda}{2\rho}\bigg) \ .
\end{align}
Now, the integrals all factorize and can be evaluated into a product of gamma functions. The result reads ($p^z=-p\cdot n_z$)
\begin{align}\label{eq:HLfn}
h_n=\frac{2(2ip^z)^{-2(n+1)\epsilon}}{(4\pi)^{\frac{D}{2}}}\frac{\pi(1-\epsilon)\Gamma\big[-(n+1)\epsilon\big]}{\sin \big[2\pi(n+1)\epsilon\big] \Gamma\big[2-(n+1)\epsilon-\epsilon\big]} \ .
\end{align}
This implies that the original diagram $V_n$ reads
\begin{align}
V_n=\frac{\alpha^{n+1}C_F}{2}\bigg(\frac{\beta_0}{2}\bigg)^n e^{(n+1)\epsilon\gamma_E}\bigg(\frac{i\sigma\mu}{2p^z}\bigg)^{2(n+1)\epsilon}\frac{f(\epsilon)^n(1-\epsilon)\Gamma\big[-(n+1)\epsilon\big]}{\epsilon^n\sin \big[2\pi(n+1)\epsilon\big] \Gamma\big[2-(n+1)\epsilon-\epsilon\big]} \ .
\end{align}
This can be re-written as
\begin{align}
&V_n=\frac{\alpha^{n+1}C_F}{2}\bigg(\frac{\beta_0}{2}\bigg)^n\frac{V(\epsilon,(n+1)\epsilon)}{(n+1)^2\epsilon^{n+2}} \ , \\
&V(\epsilon,s)=e^{s\gamma_E}\bigg(\frac{i\sigma\mu}{2p^z}\bigg)^{2s}f(\epsilon)^{\frac{s}{\epsilon}-1}\frac{(\epsilon-1)s\Gamma\big[1-s\big]}{\sin \big[2\pi s\big] \Gamma\big[2-s-\epsilon\big]} \ .
\end{align}
In particular, the bare matrix element has no factorial growth in the $n\rightarrow \infty$ limit at all. 

\subsection{Renormalization, RGE and anomalous dimensions}
Given the above, the partially renormalized bubble chain diagram reads
\begin{align}
&R_{\rm part}V_n=\frac{\alpha^{n+1}C_F}{2}\bigg(\frac{\beta_0}{2}\bigg)^n\sum_{k=0}^n(-1)^k\binom{n}{k}\frac{V(\epsilon,(n+1-k)\epsilon)}{(n+1-k)^2\epsilon^{n+2}} \  \ .
\end{align}
To proceed, one needs the formulas~\cite{Palanques-Mestre:1983ogz,Beneke:1994sw,Beneke:1995pq}
\begin{align}
   & \sum_{k=0}^n\frac{(-1)^k\binom{n}{k}}{(n-k+1)^2}=\frac{(-1)^n}{n+1}H_{n+1}  \ , \\
   & \sum_{k=0}^n\frac{(-1)^k\binom{n}{k}}{(n-k+1)}=\frac{(-1)^n}{n+1} \ , \\
   & \sum_{k=0}^{n}(-1)^k \binom{n}{k}(n-k+x)^j=0  \ , \  0\le j\le n-1  \ , \forall x \ , \label{eq:S(n,j)}\\ 
& \sum_{k=0}^n(-1)^k\binom{n}{k}(n-k+x)^n=n! \label{eq:S(n,n)} \ , \forall x \ .
\end{align}
Here, $H_n=\sum_{k=1}^n\frac{1}{k}$ is the $n$th harmonic number. 
They are all special examples of the second Stirling numbers $n!S(j+1,n+1)$ and we collect useful formulas for them in Appendix~\ref{sec:stirling}. 
Given these, the partially renormalized bubble chain reads
\begin{align}
&R_{\rm part}V_n=\nonumber \\
&\frac{\alpha^{n+1}C_F}{2}\bigg(\frac{\beta_0}{2}\bigg)^n\bigg(\frac{V(\epsilon,0)}{\epsilon^{n+2}}\frac{(-1)^n}{n+1}H_{n+1}+\frac{\frac{d}{ds}V(\epsilon,s)|_{s=0}}{\epsilon^{n+1}} \frac{(-1)^n}{n+1}+\frac{n!}{(n+2)!}\frac{d^{n+2}V(0,s)}{ds^{n+2}}|_{s=0} \bigg)\ ,
\end{align}
Now, using the regular Taylor expansion of $V(\epsilon,s)$ at $\epsilon=0$ 
\begin{align}
V(\epsilon,s)=\sum_{j=0}^{\infty}V_j(s)\epsilon^j \ ,
\end{align}
where $V_j(s)$ are regular at $s=0$, the fully renormalized result in $\overline{\rm MS}$ scheme reads 
%(no counter terms from bubble chain with $n'<n$ are needed in the large $\beta_0$ or large $n_f$ limit)
\begin{align}\label{eq:Vnfull}
&RV_n\nonumber \\ 
&=\frac{\alpha^{n+1}C_F}{2}\bigg(\frac{\beta_0}{2}\bigg)^n\bigg(\frac{(-1)^n}{n+1}H_{n+1}V_{n+2}(0)+\frac{(-1)^n}{n+1}V_{n+1}^{\prime}(0)+\frac{n!}{(n+2)!}\frac{d^{n+2}V_0(s)}{ds^{n+2}}|_{s=0} \bigg) \ .
\end{align}
Now, since $V(\epsilon,0)$ and $\frac{d}{ds}V(\epsilon,s)|_{s=0}$ are analytic functions in the neighborhood of $\epsilon=0$, the first two terms are Borel summable. On the other hand, the last term contains renormalon ambiguity due to the pole of $V_0(s)$ at $s=\frac{1}{2}$.  One can see that it is the sub-divergence subtraction that introduces the factorial growth, explaining the name for ``renormalon''. 

Given the full results for the bubble chain, as a by-product, we can check that they indeed satisfy the RGE in the large $\beta_0$ limit and extract all the anomalous dimensions. More precisely, we consider the exponentiation of the bubble chain diagrams (in the large $\beta_0$ or large $n_f$ expansion, exponentiation or not makes no difference up to ${\cal O}(\frac{1}{n_f})$)
\begin{align}
& H_{ \rm HL} = 1+\sum_{n=0}^{\infty} RV_n  \ ,  \\ 
& \hat H_{ \rm HL} = \exp \bigg[\sum_{n=0}^{\infty} RV_n \bigg] \ .
\end{align}
From the above, the scale $\mu$ dependence reads
\begin{equation}
    \frac{d \ln \hat H_{ \rm HL}}{d \ln \mu} = \frac{d H_{\rm HL}}{d \ln \mu} \ .
\end{equation}
Thus one can obtain the anomalous dimension in the large $\beta_0$ limit through calculating $\frac{d H_{\rm HL}}{d \ln \mu}$. For this purpose one needs the Taylor expansion of $V(\epsilon,s)$ at $s=0$
\begin{equation}\label{eq:Gexpand}
    V(\epsilon,s)=\sum_{j=0}^{+\infty} g_{j}(\epsilon) s^j,
\end{equation}
where $g_{j}(\epsilon)$ are analytic at $\epsilon=0$. One also needs the scale $\mu$ dependencies for the objects
\begin{align}
    &\frac{d \alpha}{d \ln \mu} = -\beta_0 \alpha^2  \ , \nonumber\\
    &\frac{d g_{0}(\epsilon)}{d \ln \mu} = 0  \ , \nonumber\\
    &\frac{d g_{1}(\epsilon)}{d \ln \mu} = 2 V_{0}(\epsilon) \ , \nonumber\\
    &\frac{d g_{n+2}(0)}{d \ln \mu} = \frac{d}{d \ln \mu} \frac{1}{(n+2)!}\frac{d^{n+2} V(0,s)}{d s^{n+2}} = 2 g_{n+1}(0) \ .
\end{align}
Based on the above identities, after certain simplifications we obtain
\begin{align}
&\frac{d H_{\rm HL}}{d \ln \mu} = \sum_{n=0}^{\infty}C_F \left( \frac{\alpha \beta_0}{2} \right)^{n} \left( \frac{ \frac{d^{n+1} V_0(\epsilon)}{d \, \epsilon^{n+1}} |_{\epsilon \rightarrow 0}}{(n+1)!}(-1)^n H_{n+1} +(-1)^n\frac{ \frac{d^{n} V_1(\epsilon)}{d \, \epsilon^{n}} |_{\epsilon \rightarrow 0}}{n!} \right) \nonumber\\
&=-\frac{2C_F}{\beta_0} \int_0^1 dt \frac{g_0\left(\frac{-\alpha \beta_0}{2} \right)-g_0\left(\frac{-\alpha \beta_0}{2} t \right)}{1-t} 
+ \alpha C_F\bigg(\tilde{g}\left( \frac{-\alpha \beta_0}{2} \right) -g_0\left( \frac{-\alpha \beta_0}{2} \right)L_z\bigg) \ .
\end{align}
where the identity $H_{n} = \int_0^1 dt \frac{1-t^n}{1-t}$ is used in the above process. The expressions for $g_0$ and $\tilde g$ reads
\begin{align}
&g_0(\epsilon) = \frac{(\epsilon -1) \Gamma (4-2 \epsilon )}{12 \pi  \Gamma (2-\epsilon )^3 \Gamma (\epsilon +1)} \ ,\nonumber\\
&\tilde g(\epsilon) = \frac{(\epsilon -1) \Gamma (4-2 \epsilon ) \left(i \pi  \epsilon \, {\rm sgn}(z) +2\gamma_E  \epsilon +\epsilon  \psi ^{(0)}(2-\epsilon )+\ln \left(6\frac{\Gamma (2-\epsilon )^2 \Gamma (\epsilon +1)}{\Gamma(4-2 \epsilon )}\right)\right)}{12 \pi  \epsilon  \Gamma (2-\epsilon )^3 \Gamma (\epsilon +1)} \ .
\end{align}
This implies that the exponentiation of the bubble chain diagram $e^{\sum_{n=0}^{\infty} RV_n}$ is closed under RGE. From the above, we can extract the cusp anomalous dimension for the bubble chain diagram~\cite{Beneke:1995pq}
\begin{align}\label{eq:cusp}
\Gamma_{\rm cusp}(\alpha) = -2 \alpha C_F g_0\left( \frac{-\alpha \beta_0}{2} \right) = \frac{\alpha  C_F \Gamma \left(\alpha  \beta _0+4\right)}{6 \pi  \Gamma \left(1-\frac{\alpha  \beta _0}{2}\right) \Gamma\left(\frac{\alpha  \beta _0}{2}+1\right) \Gamma^2 \left(\frac{\alpha  \beta _0}{2}+2\right)} \ ,
\end{align}
whose small $\alpha$ expansion (one also replaces $\beta_0 \rightarrow -\frac{n_f}{3\pi}$) reproduces the cusp anomalous dimension in the large $n_f$ limit up to NNNLO
\begin{align}
\Gamma_{\rm cusp}(\alpha) = \frac{\alpha  C_F}{\pi } -\frac{5 \alpha ^2 C_F n_f}{18 \pi ^2} - \frac{\alpha ^3 C_F n_f^2}{108 \pi ^3}  \ .
\end{align}
The result is also consistent with that from~\cite{Beneke:1995pq,Becher:2006mr} in large $n_f$ limit.
The imaginary part of the single log anomalous dimension is 
\begin{align}
{\rm Im}\left[ 2\alpha C_F\tilde{g}\left( \frac{-\alpha \beta_0}{2} \right) \right] = - \pi \Gamma_{\rm cusp} \, {\rm sgn}(z),
\end{align}
which is consistent with imaginary part of the known result. %Eq.~(\ref{eq:RGHL}).
On the other hand, the single-log anomalous dimension $\tilde \gamma_H$ for the bubble chain diagrams reads
\begin{align}
    \tilde \gamma_{H}(\alpha) =-\frac{4C_F}{\beta_0} \int_0^1 dt \frac{g_0\left(\frac{-\alpha \beta_0}{2} \right)-g_0\left(\frac{-\alpha \beta_0}{2} t \right)}{1-t} + 2\alpha C_F{\rm Re}\left[\tilde{g}\left( \frac{-\alpha \beta_0}{2} \right)\right] \ .
\end{align}
The small $\alpha$ expansion of $\tilde \gamma_H$ (one also replaces $\beta_0 \rightarrow -\frac{n_f}{3\pi}$) reads
\begin{align}
    \tilde \gamma_{H}(\alpha) = -\frac{\alpha  C_F}{\pi }+\frac{\left(80+9 \pi ^2\right) \alpha ^2 C_F n_f}{216 \pi ^2}-\frac{\alpha ^3 \left(36 \zeta_3-\frac{844}{3}+45 \pi ^2\right)C_F n_f^2}{3888 \pi ^3} \ , 
\end{align}
which reproduces the single log anomalous dimension in the large $n_f$ limit up to NNNLO.  In addition, we can also check the constant terms in $H_{HL}$ up to NNLO (setting $\beta_0 \rightarrow -\frac{n_f}{3\pi}$ and $\mu=2 p^z$)
\begin{equation}
H(L_z=0,\alpha) = 1 + \frac{\alpha  C_F}{2 \pi }\left( \frac{\pi ^2}{12}-2 \right) + \alpha^2 C_F n_f \frac{36 \zeta_3+1312+51 \pi ^2}{2592 \pi ^2} \ ,
\end{equation}
\begin{equation}
A(L_z=0,\alpha)  = - \frac{\alpha C_F}{2}  + \alpha^2 C_F n_f \left(\frac{19}{54 \pi }+\frac{\pi }{24}\right)  \ ,
\end{equation}
which are consistent with the $n_f$ part for constant terms $c_a$ and $c_H$~\cite{Ji:2023pba}.

\subsection{Borel transform and linear renormalon}
We now move to the key part of this section. Given the fully renormalized hard kernel in Eq.~(\ref{eq:Vnfull}), the Borel transform reads ($u=\frac{\beta_0t}{2}$)
\begin{align}\label{eq:BorelHLfull}
&B(u)=\sum_{n=0}^{\infty}\frac{t^n}{n!} \frac{RV_n}{\alpha^{n+1}}= R(u)+\frac{C_F}{2}\sum_{n=0}^{\infty}\frac{u^n}{(n+2)!}\frac{d^{n+2}V_0(u)}{du^{n+2}}|_{u=0}\nonumber \\
&=R(u)+\frac{C_F}{2}\frac{V_0(u)-uV_0'(0)-V_0(0)}{u^2} \ ,
\end{align}
where $R(u)$ is an entire function of $u$. Now, one has
\begin{align}\label{eq:BorelHLexp}
V_0(u)=\frac{1}{2}\frac{ue^{\frac{5u}{3}}}{u-1}\bigg(\frac{1}{\sin \pi u}+\frac{i{\rm sign}(z)}{\cos \pi u}\bigg)\left(\frac{\mu^2}{4p_z^2}\right)^u \ .
\end{align}
The Borel transform of the imaginary part reads now
\begin{align}
{\rm Im} B(u)={\rm Im} R(u)+\frac{C_F}{4 u}\bigg(\frac{e^{\left(\frac{5}{3}-L_z\right)u}}{u-1}\frac{1}{\cos \pi u}+1\bigg){\rm sign}(z) \ .
\end{align}
This implies that the Borel transform has a pole at $u=\frac{1}{2}$, with the residue 
\begin{align}\label{eq:Borel}
{\rm Im } B\left(u\rightarrow \frac{1}{2}\right)=\frac{\mu}{p^z}\frac{C_Fe^{\frac{5}{6}}}{2\pi \left(u-\frac{1}{2}\right)}{\rm sign}(z) \ .
\end{align}
Clearly, the linear renormalon appears only in the imaginary part.  As such, the asymptotic behavior of the phase angle $A(L_z,\alpha)$ can be predicted as
%($b=\frac{\beta_1}{2\beta_0^2}$)
%\begin{align}\label{eq:asympA}
%A^{(n)}\bigg(L_z,\alpha(\mu)\bigg) \rightarrow  \tilde N_m \frac{\mu}{p_z}\left(\frac{\beta_0}{2\pi}\right)^n\frac{\Gamma(n+1+b)}{\Gamma(b+1)} \ .
%\end{align}
%In the $\overline {\rm MS}$ scheme, one has
%\begin{align}\label{eq:amplitudere}
%\tilde N_m=-\frac{C_Fe^{\frac{5}{6}}}{\pi} \ .
%\end{align}
\begin{align}
A_{\rm asym}\bigg(L_z,\alpha(\mu)\bigg) = -2\frac{\mu}{p^z} \sum_{n=0} r_{n} \alpha^{n+1},
\end{align}
where the factor 2 here comes from the factor 2 in the definition Eq.~(\ref{eq:defJf}). The coefficient is
\begin{align}\label{eq:asympA}
r_{n} = N_m \beta_0^n \Gamma(n+1).
\end{align}
In the $\overline {\rm MS}$ scheme, through comparing Eqs.~(\ref{eq:Borel})~(\ref{eq:asympA}), one has 
\begin{align}\label{eq:amplitudere}
N_m= \frac{C_Fe^{\frac{5}{6}}}{\pi} = 0.977 \ .
\end{align}
It is interesting to notice that the above amplitude of the renormalon ambiguity is exactly the same as that for the linear divergence or the heavy-quark pole mass in the large $\beta_0$ limit~\cite{Beneke:1994rs,Pineda:2001zq,Ayala:2014yxa}. Eq.~(\ref{eq:Borel}),(\ref{eq:asympA}) and (\ref{eq:amplitudere}) are the main results of this section. 

Through requiring that the renormalon ambiguity is renormalization scale independent, one can generalize Eq.~(\ref{eq:asympA}) beyond large $\beta_0$ limit~\cite{Beneke:1994rs,Pineda:2001zq},
\begin{align}\label{eq:asympAb}
r_{n} = N_m \beta_0^n \frac{\Gamma(n+1+b)}{\Gamma(1+b)} \left( 1+ \frac{b}{b+n} c_1 + \frac{b(b-1)}{(n+b)(n+b-1)} c_2 + ...\right),
\end{align}
where $b = \frac{\beta_1}{\beta_0^2}$, $c_1 = \frac{\beta_1^2 - \beta_0 \beta_2}{b \beta_0^4}$ and $c_2 = \frac{-\beta _3 \beta _0^4+2 \beta _1 \beta _2 \beta _0^3+\left(\beta _2^2-\beta _1^3\right)\beta _0^2-2 \beta _1^2 \beta _2 \beta _0+\beta _1^4}{2 (b-1) b \beta _0^8}$. Our convention for the beta function is $\frac{d \alpha}{d \ln \mu} = -\beta_0 \alpha^2 -\beta_1 \alpha^3 - \beta_2 \alpha^4 -...$ Here we propose a conjecture that the amplitude $N_m$ for the phase angle $A$ is the same as that for the linear divergence or the heavy-quark pole mass even beyond large $\beta_0$ limit, which means 
\begin{align}
    N_m = 
    \begin{cases}
        0.622 & n_f = 0 \\
        0.609 & n_f = 1 \\
        0.593 & n_f = 2 \\
        0.575 & n_f = 3 \\
        0.552 & n_f = 4 \\
        0.524 & n_f = 5 
    \end{cases}
\end{align}
where those values are obtained based on the method in Ref.~\cite{Pineda:2001zq} using the perturbation series between the $\overline{\rm MS}$ and the pole quark masses~\cite{Melnikov:2000qh}. Here the $N_m=0.622$ for $n_f=0$ is consistent with that obtained from the linear divergence series $N_m = 0.660(56)$~\cite{Bali:2013pla}. There are several evidences for this conjecture. First, they share the same $N_m$ in the large $\beta_0$ limit. Second, as we will discuss later, the renormalon series for the heavy light Sudakov hard kernel matches that of NLP TMDWF, whose Borel transformation is very similar to that of the static potential, see Eq.~(\ref{eq:NLPresults}). As shown in~\cite{Pineda:2002se}, the renormalon series of the static potential cancels that of the self energy in a Wilson loop.

\section{Renormalon cancellation and power corrections to threshold asymptotics of space-like quark-bilinear coefficient functions} \label{sec:threshold}
In this section, we investigate the role played by the linear renormalon of $H_{\rm HL}$ in an important place where it appears, the threshold limit of the hard function or coefficient function that relates the space-like quark-bilinear matrix elements (``qPDF'' or ``pPDF'') to twist-two light-front correlators.  Throughout this section, $|z|\Lambda_{\rm QCD} \ll 1$ is always understood. 

The main object is the perturbative quark correlator in coordinate space
 \begin{align}\label{eq:deftildef}
{\cal H}\bigg(z^2,\lambda,\alpha(\mu)\bigg)=\langle p |\bar \psi(z)\gamma^z[zn_z,0]\psi(0)  | p \rangle \ ,
\end{align}
where $|p\rangle$ is the on-shell external quark state with $p^z>0$ and $\lambda=z p^z$. All the IR divergences are subtracted in the ${\overline MS}$ scheme through convolution and the tree level is normalized to $e^{-i\lambda}$ . In this paper we only consider quark non-singlet  un-polarized contributions with $\vec{p}_\perp \equiv 0$. Clearly, this correlator is the leading-twist OPE coefficient function for the space-like quark-bilinear transformed into coordinate space.  Our convention for the moment space read
\begin{align}
{\cal H}(z^2,\lambda,\alpha(\mu))=\int_{-1}^{1} d\alpha e^{-i\lambda \alpha}{\cal H}(z^2,\alpha,\alpha(\mu)) \ , \\ 
{\cal H}_N(z^2,\alpha(\mu))=\int_{-1}^1 d\alpha \alpha^N  {\cal H}(z^2,\alpha,\alpha(\mu)) \ .
\end{align}
As usual, the support in $\alpha$ is in $[-1,1]$ (this implies that the coordinate space version is analytic in the whole complex $\lambda$ plane) and for the purpose of extracting the $\alpha \rightarrow 1$ threshold asymptotics through the $N\rightarrow \infty$ limit,  in principle one should integrate from $0$ to $1$. However, due to lack of contributions at ${\cal O}(1+\alpha)^{-1}$ as $\alpha \rightarrow -1$ for quark non-singlet un-polarized coefficient functions, to leading power in $\frac{1}{N}$, the $[-1,0]$ part can be added as well. Also notice that at the level of bubble chain diagrams, supports are always in $[0,1]$.

As shown in~\cite{Ji:2023pba} to NNLO, the coordinate space coefficient function in the {\it threshold limit} $\lambda \rightarrow \infty$\footnote{Here we should note that in the large $\lambda$ limit, there are contributions without the $e^{-i\lambda}$ factor as well, see Eq.~(\ref{eq:fullasympthre}) below. These contributions are non-threshold and should not be considered. At leading and next-to-leading powers in $\frac{1}{\lambda}$, these non-threshold contributions do not appear. An equivalent way to define the threshold limit is to use $\lambda=\lambda+i|\lambda|0$ with infinitely-small positive imaginary parts or simply use $\lambda \rightarrow +i\infty$. In this way threshold contributions always dominate. }, has the following factorization formula at leading {\it threshold power} in $\frac{1}{|\lambda|}$ 
\begin{align}\label{eq:threshfac}
{\cal H} \bigg(z^2,\lambda,\alpha(\mu)\bigg)=e^{-i\lambda}H^2_{\rm HL}\bigg(\ln \frac{-2i\lambda}{|z|\mu},\alpha(\mu)\bigg)J\bigg(l_z,\alpha(\mu )\bigg) \ .
\end{align}
Here $H_{\rm HL}$ is nothing but the HL Sudakov hard kernel, and the LP space-like threshold soft factor $J$ or ``jet function'' is a specific Wilson-loop average involving space-like and light-like gauge-links defined in~\cite{Ji:2023pba}
\begin{align}\label{eq:defJ}
J(l_z,\alpha(\mu))=\langle \Omega|{\cal T}[zn_z+\infty n^+,zn_z][zn_z,0][0,-\infty n^+]|\Omega\rangle \ ,
\end{align}
where $l_z \equiv \ln \frac{e^{\gamma_E}|z|\mu}{2}$ is the natural logarithm in $z$. Here we emphasize that in the threshold $\lambda \rightarrow \infty$ limit, the threshold logarithms are due to the {\it hard scale} $\frac{|\lambda|}{|z|} \gg \frac{1}{|z|} \gg \Lambda_{\rm QCD}$ and are exactly resummed using the heavy-light Sudakov hard kernel. Notice that the threshold power is finer the standard ``twist-power'' counted in $z\Lambda_{\rm QCD}$, it is counted in $\frac{1}{\lambda}$ or $\frac{1}{N}$. A single twist in the threshold limit can split into infinitely many threshold powers and there is no conflict between these two expansions due to the fact that the threshold limit is UV in our case. 

Although the initial work is in coordinate space, we can convert all the results to the moment space.  More precisely, to leading threshold power in $\frac{1}{N}$, one has for $N$-th moment for the  coefficient function
\begin{align}\label{eq:thresholdmomemtfac}
{\cal H}_N(z^2,\alpha(\mu))|_{N\rightarrow \infty} \rightarrow  H^2_{\rm HL} \bigg( \ln \frac{2N}{\mu|z|},\alpha(\mu)\bigg) J\bigg(l_z ,\alpha(\mu )\bigg)\bigg(1+{\cal O}\bigg(\frac{1}{N}\bigg) \bigg)\ ,
\end{align}
which can be converted into the less convenient ``A-B-C'' form.  Again, the threshold scale $\frac{2N}{|z|}$ in moment space is hard.  To obtain the fully resumed form, one needs the RGEs 
\begin{align}
&2\mu \frac{d}{d\mu} \ln H_{\rm HL} \bigg( \ln \frac{2N}{\mu|z|},\alpha(\mu)\bigg) =2\Gamma_{\rm cusp}\ln \frac{2N}{\mu|z|}+\tilde \gamma_H \ , \\ 
&\mu \frac{d}{d\mu} \ln J\bigg(l_z,\alpha(\mu)\bigg)=2\Gamma_{\rm cusp}\ln \frac{e^{\gamma_E}\mu|z|}{2}-\tilde \gamma_J \ .
\end{align}
In the above, $\tilde \gamma_J$ is the single-log anomalous dimension for the space-like threshold soft factor and equals to $2\gamma_{HL}-2\gamma_s$. The RGEs of the hard kernel and the threshold soft factor allow the systematic resummation of threshold-logarithms in a way similar to Eq.~(5.5) of~\cite{Beneke:1995pq} without using the ``A-B-C'' form. Moreover, after partial integration, all the $\alpha(\mu)$ dependencies can be factorized out completely and the controlled asymptotic expansion in the threshold limit in moment space takes a simple factorized form in terms of $\alpha(z_N)\equiv \alpha \left(\frac{2N}{|z|}\right)$ and $\alpha(z) \equiv \alpha\left(\frac{2e^{-\gamma_E}}{|z|}\right)$ as we show now. The controlled expansion in the coordinate space is similar.

To write out the controlled expansion in the $N\rightarrow \infty$ limit for the ${\cal H}_N$, it is convenient to define the following regular functions ($\beta(\alpha)=-\beta_0\alpha^2-\beta_1\alpha^3+..$, $\Gamma_{\rm cusp}(\alpha)=\gamma_0\alpha+\gamma_1\alpha^2+..$)
\begin{align}
&f_{\Gamma}(\alpha)=\int_{0}^{\alpha}d\alpha'\bigg(\frac{\Gamma_{\rm cusp}(\alpha')}{\beta(\alpha')}+\frac{\gamma_0}{\beta_0 \alpha'}\bigg)=f_{\Gamma}^{0}\alpha+f_{\Gamma}^{(1)}\alpha^2+.... \ ,  \\ 
&g_{\Gamma}(\alpha)=\int_{0}^{\alpha}d\alpha'\bigg(\frac{f_{\Gamma}(\alpha')}{\beta(\alpha')}+\frac{f_{\Gamma}^0}{\beta_0 \alpha'}\bigg)= g_{\Gamma}^{0}\alpha+g_{\Gamma}^{(1)}\alpha^2+...
\end{align}
and 
\begin{align}
&k_{\beta}(\alpha)=\int_{0}^{\alpha}d\alpha' \bigg(\frac{1}{\beta(\alpha')}+\frac{1}{\beta_0 \alpha'^2}-\frac{\beta_1}{\beta_0^2 \alpha'}\bigg)=k_{\beta}^0\alpha+k_{\beta}^1\alpha^2+...  \ , \nonumber \\ 
&l_{\beta}(\alpha)=
-\int_{0}^{\alpha}\frac{d\alpha'}{\alpha'} \int_{0}^{\alpha'} \bigg(\frac{1}{\beta(\alpha'')}+\frac{1}{\beta_0 (\alpha'')^2}-\frac{\beta_1}{\beta_0^2 \alpha''}\bigg) d\alpha''=l_{\beta}^{0}\alpha+l_{\beta}^{(1)}\alpha^2+... \ ,
\end{align}
The definition of $f_{\gamma}$ is similar for any anomalous dimension $\gamma$. They are all renormalon-free provided that the corresponding $\beta$ function and anomalous dimensions are renormalon-free. In terms of the above, one has the fully-resummed form in the Mellin space to leading power in $\frac{1}{N}$ :
\begin{bbox}
\begin{align}\label{eq:thresholdcon}
&{\cal H}_N(z^2,\alpha(\mu))=  H_{\rm HL}^2\bigg(0,\alpha(z_N)\bigg)J(0,\alpha(z))\exp \bigg(\hat O_H(\alpha(z_N))+\hat O_J(\alpha(z))\bigg) \nonumber \\ 
& \times \exp \bigg(2\ln Ne^{\gamma_E}\bigg(f_{\Gamma}(\alpha(\mu))-\frac{\gamma_0}{\beta_0}\ln \alpha(\mu)\bigg)+f_{s+V+2F}(\alpha(\mu))-\frac{\gamma^0_{s+V+2F}}
{\beta_0}\ln \alpha(\mu)\bigg)  \ .
\end{align}
\end{bbox} 
In the above, the $\mu$ independent re-summation factors $\hat O_H$ and $\hat O_J$ are naturally written in terms of $\alpha(z_N)\equiv \alpha \left(\frac{2N}{|z|}\right)$ and $\alpha(z) \equiv \alpha\left(\frac{2e^{-\gamma_E}}{|z|}\right)$ as
\begin{bbox}
\begin{align}\label{eq:resummationfactor}
&\hat O_H(\alpha(z_N))=-\frac{2\gamma_0}{\beta_0^2 \alpha(z_N)}\ln \frac{1}{e\alpha(z_N)}+\frac{\beta_1\gamma_0}{\beta_0^3} \ln^2 \alpha(z_N) +\frac{2f_{\Gamma}^0+\gamma_H^0}{\beta_0}\ln \alpha(z_N)\nonumber \\
&+\frac{2\gamma_0}{\beta_0}\ln \alpha(z_N)k_{\beta}(\alpha(z_N))+\frac{2\gamma_0}{\beta_0}l_{\beta}(\alpha(z_N))-2g_{\Gamma}(\alpha(z_N))-f_{H}(\alpha(z_N)) \ ,  \nonumber \\ 
&\hat O_J (\alpha(z))=\frac{2\gamma_0}{\beta_0^2 \alpha(z)}\ln \frac{1}{e\alpha(z)}-\frac{\beta_1\gamma_0}{\beta_0^3} \ln^2 \alpha(z)-\frac{2f_{\Gamma}^0+\gamma_{J}^0}{\beta_0} \ln \alpha(z)\nonumber \\ 
&-\frac{2\gamma_0}{\beta_0} \ln \alpha(z) k_{\beta}(\alpha(z))-\frac{2\gamma_0}{\beta_0}l_{\beta}(\alpha(z)) +2g_{\Gamma}(\alpha(z))+f_{J}(\alpha(z)) \ .
\end{align}
\end{bbox}
Notice that the $\alpha(z_N)$ factorizes completely from $\alpha(z)$, explaining the name of threshold factorization. Moreover, in the $N\rightarrow \infty$ limit, the re-summation factors $\hat O_H$  is not only a controlled asymptotic expansion, but also free from any renormalon. Nevertheless, the {\it initial condition} $H_{HL}^2(0,\alpha(z_N))$ of the 
heavy-light hard kernel, although still a controlled asymptotic expansion, has linear renormalon ambiguity of the order 
\begin{align}
\frac{|z|\Lambda_{\rm QCD}}{N} \ll |z|\Lambda_{\rm QCD} ,\frac{1}{N} \ll 1 \ .
\end{align}
This reflects the intrinsic order of ambiguity for the separation between the leading and the next-to-leading threshold powers or equivalently, the separation between two hard scales $\frac{|\lambda|}{|z|}$ and $\frac{1}{|z|}$. Notice that unlike the ``standard'' case of the DY or DIS threshold limits, the threshold expansion here do not compete with the standard twist expansion due to the fact that $N$ is in the denominator instead of numerator. Indeed,  higher-order renormalons in the HL hard kernel correspond to $\frac{(z\Lambda_{\rm QCD})^k}{N^k}, k\ge 2$ ambiguities and are always controlled in the large $N$ limit, see Eq.~(\ref{eq:threshmoment}). Clearly, this is due to the fact that the threshold scale $z_N$ for the space-like parton distribution is hard and the $N\rightarrow \infty$ limit is a UV limit.

To demonstrate the above statements concerning the structure of renormalon, in this section we investigate the bubble chain diagram for the coefficient function ${\cal H}$ in the threshold limit. The purpose is to show
\begin{enumerate}
    \item At the level of bubble chain diagram, the threshold factorization formula is correct to leading power in $\frac{1}{\lambda}$ or $\frac{1}{N}$ (but to all logarithmic orders). More precisely, we perform the threshold expansion of the bubble chain diagram to leading power both in coordinate and moment space. We show that the resulting hard kernel exactly reproduces that calculated in Sec.~\ref{sec:HLbubble}, and the resulting ``jet  function'' exactly reproduces the bubble chain diagrams for the space-like threshold soft factor. 
    \item Given the above, the threshold factorization formula is established at the level of bubble chain diagram. We further investigate the role played by the linear renormalon in the heavy-light hard kernel $H_{\rm HL}$. We show that the linear renormalon cancels between the leading and next-to-leading threshold powers by expanding the full bubble chain results to NLP.   We further identify the operator definition of the obtained NLP contribution (at ${\cal O}(\frac{1}{|\lambda|})$) and show that the $s=\frac{1}{2}$ renormalon for the NLP soft factor is UV in nature and is due to implicit ambiguity of subtracting UV divergences. Notice this has nothing to do with the linear UV divergence for space-like gauge-link self-interactions.
\end{enumerate}

To summarize, the factorization formula Eq.~(\ref{eq:thresholdcon}) can be cast into a controlled asymptotic expansion in the threshold limit using the RGEs.  Nevertheless, the separation between leading and next-to-leading threshold power contributions (more precisely, separation between $\frac{|\lambda|}{|z|}$ and $\frac{1}{|z|}$) requires a scheme, reflected in the (IR) renormalon of the leading threshold power hard contribution of order $\frac{\Lambda_{\rm QCD}}{\frac{2N}{|z|}}=\frac{|z|\Lambda_{\rm QCD}}{2N}$ and (UV) renormalon of the next-to-leading threshold power ``soft'' contribution of the same order ($\frac{1}{N} \times \Lambda_{\rm QCD}|z|$). The scheme dependency  cancels between LP and NLP, or equivalently, the renormalon cancels between LP and NLP.  

It is also helpful to make the following comparisons to the DY threshold limit investigated in~\cite{Beneke:1995pq}:
\begin{enumerate}
    \item In the DIS or DY case\footnote{Notice that if one define the DIS-like structure function $F(z^2,\lambda)$ in {\it coordinate space}, then the threshold limit $\lambda \rightarrow \infty$ for the coefficient function $F_q(z^2,\lambda)$ or $N\rightarrow \infty$ for $F_{q,N}(z^2)$ become UV as well. On the other hand, in this case the hard scale is $\frac{\lambda}{z^2}$ instead of $\frac{\lambda^2}{z^2}$ and the leading renormalon for the Sudakov-like hard kernel is quadratic.}, in the $z \rightarrow 1$ limit the threshold scales $Q(1-z)$ or $Q\sqrt{1-z}$ are {\it soft}. Therefore, the resummation factors in the ``A-B-C'' form are sensitive to the Landau pole and finite-order truncation of the anomalous dimensions can create renormalons. In our case, in the $\lambda \rightarrow \infty$ or $N \rightarrow$ limit the threshold scales $\frac{\lambda}{|z|}$ or $\frac{N}{|z|}$ are UV and the resummation factors are always far away from Landau pole. Moreover, we use the RGEs for the hard kernel and the soft factor to perform the resummation in a way similar to Eq.~(5.5) of~\cite{Beneke:1995pq},  instead of truncating the ``A-B-C''.  As such, the linear renormalon observed here has absolutely no relation to resummation. 
    \item Due to the above, in~\cite{Beneke:1995pq} the authors focus on soft and collinear contributions in the threshold limit. They show that the ``soft-collinear'' contribution which dominates the DLA cancels with ``wide-angle'' soft emissions in terms of linear renormalon. In particular, they show that the bubble chain diagram for the DY threshold soft factor is free from linear renormalon and this is consistent with the power correction extracted using finite gluon mass. In our case, the similar object that captures ``IR'' contributions is the space-like threshold soft factor $J(l_z,\alpha(\mu))$ (similar to the DY case, it is also a Wilson-loop) at scale $\frac{1}{|z|}$ and it is free from linear renormalon as well (except the overall UV renormalon for the space-like gauge-link inherited from the definition of pPDF and have nothing to do with threshold expansion). But this is not the object that is responsible for the threshold logarithms in our case.
    \item In our case, the object that is responsible for the threshold logarithms is in fact the heavy-light hard  kernel. It is purely virtual instead of real, and it is not purely eikonal. Unlike the standard light-light Sudakov hard kernel (namely, the $H(\alpha(Q))$ in Eq.~(1.4) of~\cite{Beneke:1995pq} ), in our hard kernel, the heavy gauge-link and the light parton are not symmetric, as a result, when a finite gluon mass is added, it develops linear power correction, consistent with the observed linear renormalon. This also means that the cancellation of the linear renormalon must involve sub-eikonal effects, therefore must be between leading and next to leading threshold powers, consistent with our observation.  
\end{enumerate}

Here we also comment in terms of the relation to the original calculation in Ref.~\cite{Braun:2018brg}. In fact, the only question we need to answer is that: why in the original work there is almost no linear renormalon (except for the trivial one localized into the gauge-link self energy diagrams), but in this work there is. The reason is again due to the projection to the leading threshold power. Before the threshold expansion, both the virtual and the real diagrams are of the same twist in $|z|\Lambda_{\rm QCD}$ and the linear renormalon in the virtual diagram is completely cancelled by the real diagram. After projecting to the leading threshold power, the virtual diagrams survive, but the real diagrams eikonalize and become part of the space-like threshold soft factor. The contributions in the real diagrams that cancel the linear renormalon are sub-eikonal and move to the next-to-leading threshold power $\frac{1}{N}$. The linear renormalon then simply measures the intrinsic ambiguity of the threshold-power separation (but not $z,\Lambda_{\rm QCD}$ separation).

Let's emphasize here that the order of our linear renormalon ambiguity is $\frac{|z|\Lambda_{\rm QCD}}{N}$ and is much smaller in the $N\rightarrow \infty$ limit than the ``genuine'' high-twist contribution ${\cal O}(|z|^2\Lambda^2_{\rm QCD})$. In particular, the threshold expansion is still performed within a given twist, including the order of ambiguities. Also notice that although we call the linear renormalon in the HL hard kernel ``IR'', it is only IR with respect to the scale $\frac{|\lambda|}{|z|}$ for the hard kernel, but not to $\frac{1}{|z|}$. In fact, we will see that the linear ``IR'' renormalon cancels with UV renormalon of NLP ``soft'' contributions at scale $\frac{1}{|z|}$. 

Due to the above, the linear renormalon should not be regarded as a prediction of presence of genuine non-perturbative effects enhanced in the threshold limit. Instead, it should be interpreted as predicting the existence of next-to-leading threshold power contributions to the leading twist coefficient function that has linear UV renormalon.  Moreover, all the ``threshold renormalons'' generated after threshold expansion always cancel between different threshold powers,  and the ``twist-expansion renormalons'' separating the leading and high powers in the $z\Lambda_{\rm QCD}$ expansion remain the same as given by Eq.~(48) and Eq.~(49) of the original work~\cite{Braun:2018brg}. In the threshold region, they start contributing only from $\frac{(z\Lambda_{\rm QCD})^2}{N}$ at $\frac{1}{N}$ and $(z\Lambda_{\rm QCD})^4$ at $N^0$. Finally, both the ``threshold renormalons'' and the ``genuine'' renormalons in~\cite{Braun:2018brg} never increases as $N\rightarrow \infty$. Genuine non-perturbative effects are uniformly bounded by $(z\Lambda_{\rm QCD})^2$ itself in an $N$-independent way.

\subsection{Collection of full results before threshold limit}
In this section we consider the bubble chain diagrams for Eq.~(\ref{eq:deftildef}) between the incoming quark and the space-like gauge-link corresponding to Fig.~1a in~\cite{Braun:2018brg} (another diagram Fig.~1b that couples the outgoing quark and the gauge link contributes equally). Only these two diagrams contribute to non-trivial threshold logarithms in the threshold limit. The gauge-link self-interaction diagram Fig.~1d depends only on $z^2$ and is part of the LP threshold soft factor. The one gluon exchange diagrams Fig.~1c between quark and anti-quark start contributing only at $\frac{1}{N^2}$ and contain no linear renormalons either before or after threshold expansion so will not be considered. Notice that all results are normalized in a way such that the tree-level diagram reads $e^{-i\lambda}$ and only un-polarized component is considered. In the notation of~\cite{Braun:2018brg}, our results always corresponds to $H^{\parallel}$, although for the diagrams we considered there is no difference between $H^{\parallel}$ and $H^{\perp}$.

Since the calculation for all the diagrams has already been performed in~\cite{Braun:2018brg}, we will not present calculation details. Instead, we start from the following expression for un-renormalized $n$-bubble chain contribution (corresponding to Fig.~1a in~\cite{Braun:2018brg})
\begin{align}
{\cal H}_{n}(z^2,\lambda)=g^2C_F\alpha(\mu)^n\bigg(\frac{\beta_0}{2}\bigg)^n\mu_0^{2\epsilon}\mu^{2n\epsilon}e^{n\epsilon \gamma_E}\frac{1}{\epsilon^n}f(\epsilon)^n \mathfrak{h}_n(z^2,\lambda) \ ,
\end{align}
one has
\begin{align}
&\mathfrak{h}_n(z^2,\lambda)=\frac{1}{(4\pi)^{\frac{D}{2}}\Gamma(1+n\epsilon)}\nonumber \\ 
&\times\int_{0}^{1}dt \int_{0}^{\infty} d\rho \rho^{-1+(n+1)\epsilon}\int_0^1dx \bar x^{n\epsilon}e^{-\frac{z^2t^2}{4\rho}-i\lambda(1- x t)} \bigg(2(1-x)i\lambda+\frac{z^2t}{2\rho}\bigg) \ .
\end{align}
After integrating over $x$ and $t$, the above leads to ( $\, _2\tilde{F}_2$ is the regularized hypergeometric function and $s=(n+1)\epsilon$) 
\begin{align}\label{eq:bubblecoord}
&\mathfrak{h}_n(z^2,i\lambda)=e^{\lambda}\frac{2\Gamma(1-s)\Gamma(2s)}{(4\pi)^{\frac{D}{2}}}\left(\frac{z^2}{4}\right)^s\bigg( \, _2\tilde{F}_2(1,2 s;-\epsilon+s+2,2 s+1;-\lambda)\nonumber \\ 
&+2 \lambda \left(\, _2\tilde{F}_2(1,2 s+1;-\epsilon+s+2,2 s+2;-\lambda)-\, _2\tilde{F}_2(2,2 s+1;-\epsilon+s+3,2 s+2;-\lambda)\right)\bigg) \ .
\end{align}
Notice that the above is equivalent to Eq.~(A6) in~\cite{Braun:2018brg} and we will come back to this in Sec.~\ref{sec:resurgent}. Converting into the moment space, one has for the moments (the notation for $\mathfrak{h}_{n,N}(z^2)$ follows the same convention as $\mathfrak{h}_n(z^2,\lambda)$)
\begin{align}\label{eq:bubblemoment}
\mathfrak{h}_{n,N}(z^2)=-\frac{2\Gamma(-s)}{(4\pi)^{\frac{D}{2}}\Gamma(s-\epsilon+1)}\bigg(\frac{z^2}{4}\bigg)^s\bigg(NI(2s,s-\epsilon+1;N-1)+sI(2s-1,s-\epsilon;N)\bigg) \ .
\end{align} 
Here we adopted the notation  
\begin{align}
&I(a,b;N)\equiv \int_{0}^1 t^a dt\int_{0}^1 x^b \left(1-\bar x t\right)^Ndx
\nonumber \\ 
&=-\Gamma(N+1)\Gamma (-a)\left( \frac{\, _3F_2(1,-a,-b;1-a,N+2;1)}{\Gamma(1-a)\Gamma(N+2)}-\frac{\Gamma(a+1) \Gamma (b+1)}{\Gamma (-a+b+1) \Gamma (a+N+2)}\right) \ .
\end{align}
The above serve as the starting point of the following discussions. Notice that the above has no pole at $s=\frac{1}{2}$, $\epsilon=0$ , thus after renormalization in standard manner, no linear renormalon can be generated. This is consistent with the original result in~\cite{Braun:2018brg}. 

\subsection{Threshold limit in coordinate and moment space}
Now we study the threshold limit of the full results above. We perform large $\lambda$ and large $N$ expansions up to NLP. We show that the threshold logs at leading power are due to hard scale $\frac{\lambda^2}{z^2}$ and the
corresponding hard contributions factorize exactly into the heavy-light hard kernel calculated in Sec.~\ref{sec:HLbubble}. The soft contribution at LP corresponds exactly to the bubble chain diagram between $n^+$ and $n^z$ for the space-like threshold soft factor Eq.~(\ref{eq:defJ}). This verifies the threshold factorization formulas Eq.~~(\ref{eq:threshfac}) and Eq.~(\ref{eq:thresholdmomemtfac})  at the level of bubble chain diagram (previously it has been verified to NNLO). The pole at $s=\frac{1}{2}$ cancel between LP hard and NLP soft contributions.

We first consider the threshold limit in the coordinate space. Given Eq.~(\ref{eq:bubblecoord}), its asymptotic expansion in large $\lambda$ limit reads
\begin{align}\label{eq:thresholdlimitc}
&\mathfrak{h}_n(z^2,i\lambda)=\frac{2\pi e^{\lambda}}{(4\pi)^{\frac{D}{2}}}\left(\frac{|z|}{2\lambda}\right)^{2s}\frac{(1-\epsilon)\Gamma(-s)}{\sin 2\pi s \Gamma(2-s-\epsilon)}\nonumber \\ 
&-\frac{e^{\lambda}}{(4\pi)^{\frac{D}{2}}}\bigg(\frac{z^2}{4}\bigg)^s\bigg(\frac{\Gamma(-s)}{s\Gamma(1-\epsilon+s)}+\frac{2(\epsilon-1)\Gamma(-s)}{(2s-1)\Gamma(1-\epsilon+s)\lambda}+{\cal O}\bigg(\frac{1}{\lambda^2}\bigg)\bigg) \ .
\end{align}
It is clear now that the first term, after renormalization, simply generates all the threshold logarithms at leading power. From the combination $\left(\frac{|z|}{2\lambda}\right)^{2s}$ it is clear that threshold logs are due to the hard scale $\frac{4\lambda^2}{z^2}$ and is UV in nature. Moreover, this term is nothing but the $n$-bubble diagram for the heavy-light hard kernel given in Eq.~(\ref{eq:HLfn}). On the other hand, it is not hard to show that the first term in the second line simply reproduces the $n$-bubble diagram between $n^+$ and $n_z$ for the LP space-like threshold soft factor in Eq.~(\ref{eq:defJ}) and is free from singularity at $s=\frac{1}{2}$. Therefore, the leading power threshold factorization formula Eq.~(\ref{eq:threshfac}) is verified to all orders at the level of bubble chain diagrams. Moreover, the $s=\frac{1}{2}$ pole for the HL hard kernel is simply cancelled by the NLP contribution 
\begin{align}\label{eq:thresholdNLPcoor}
\mathfrak{h}_n^{\rm NLP}(z^2,i\lambda)=-\frac{ e^{\lambda}}{\lambda}\frac{2}{(4\pi)^{\frac{D}{2}}}\left(\frac{z^2}{4}\right)^s\frac{(\epsilon-1)\Gamma(-s)}{(2s-1)\Gamma(1-\epsilon+s)} \ .
\end{align}
We will show that the $s=\frac{1}{2}$ pole for the NLP contribution is in fact UV in nature, in subsection.~\ref{sec: UVrenor}.

Similarly, starting from Eq.~(\ref{eq:bubblemoment}) one can perform the threshold limit $N\rightarrow \infty$ in the moment space as well. The result reads
\begin{align}\label{eq:thresholdlimitm}
\mathfrak{h}_{n,N}(z^2)=&\frac{2\pi}{(4\pi)^{\frac{D}{2}}}\left(\frac{z^2}{4N^2}\right)^s\frac{(1-\epsilon)\Gamma(-s)}{\sin 2\pi s \Gamma(2-s-\epsilon)}\bigg(1-\frac{s(2s+1)}{N}+{\cal O}\bigg(\frac{1}{N^2}\bigg)\bigg) \nonumber \\ 
&-\frac{1}{(4\pi)^{\frac{D}{2}}}\bigg(\frac{z^2}{4}\bigg)^s\bigg(\frac{\Gamma(-s)}{s\Gamma(1-\epsilon+s)}+\frac{2(\epsilon-1)\Gamma(-s)}{(2s-1)\Gamma(1-\epsilon+s)N}+{\cal O}\bigg(\frac{1}{N^2}\bigg)\bigg) \ .
\end{align}
The first line again corresponds to the heavy-light hard kernel, while the second line corresponds to soft factors at LP and NLP. Therefore, the threshold factorization formula Eq.~(\ref{eq:thresholdmomemtfac}) is established at LP in the moment space as well. Notice that the term $-\frac{s(2s+1)}{N}$ in the first line is due to ``kinematic'' $\frac{1}{N}$ contributions and is absent in the coordinate space version Eq.~(\ref{eq:thresholdlimitc}). They can be generated naturally when converting the coordinate space version to moment space. The full contribution to the moment from the hard kernel simply reads
\begin{align}\label{eq:threshmoment}
\mathfrak{h}_{n,N}^h(z^2)=\frac{2\pi}{(4\pi)^{\frac{D}{2}}}\left(\frac{z^2}{4}\right)^s\frac{(1-\epsilon)\Gamma(-s)}{\sin 2\pi s \Gamma(2-s-\epsilon)}\frac{\Gamma(N+1)}{\Gamma(N+1+2s)} \ ,
\end{align}
which reproduces the first line in Eq.~(\ref{eq:thresholdlimitm}) after expanding in $N$. Notice the above is similar to Eq.~(5.19) of~\cite{Beneke:1995pq}, in fact in our notation Ref.~\cite{Beneke:1995pq}  has $\left(\frac{1}{Q^2}\right)^s\frac{\Gamma(N+1)}{\Gamma(N+1-2s)}$ . The difference in sign of the $2s$ in the denominator reflects the deep fact that the threshold scale in~\cite{Beneke:1995pq} is $\frac{Q^2}{N^2} \ll Q^2$ while in our case is $\frac{N^2}{z^2} \gg \frac{1}{z^2}$\footnote{The UV nature of the (fixed $z^2$)  $N\rightarrow \infty$ limit  has also been noticed in Ref.~\cite{Gao:2021hxl} using other arguments. }.  Also notice that to Fourier transform the NLP soft contribution, one must define $\frac{1}{\lambda}$ as the  boundary value approached from upper half-plane
\begin{align}
\lim_{\eta \rightarrow 0^+}\frac{1}{\lambda+i\eta}={\cal P}\frac{1}{\lambda}-i\pi \delta(\lambda) \ .
\end{align}
In this way one maintains the correct support property and one can show that the result Eq.~(\ref{eq:thresholdlimitc}) after converting to the moment space (regarding the issue of support in $\alpha$ space, see Sec.~\ref{sec:resurgent}), simply reproduces Eq.~(\ref{eq:thresholdlimitm}).

To summarize, the threshold factorization formulas Eq.~~(\ref{eq:threshfac}) and Eq.~(\ref{eq:thresholdmomemtfac}) are verified exactly at the level of un-renormalized $n$-bubble chain level. After renormalization one obtains the corresponding renormalized version. Notice that the $\frac{1}{s^2}$ poles at LP simply cancel between the hard and soft contributions, and the total anomalous dimension remains single-log in nature and simply compensates the $N\rightarrow \infty$ limit of the $\gamma^{qq}_N$ up to $2\gamma_F$
\begin{align}
\int_{0}^1 dz z^N P_{qq}(z)=\gamma_N \rightarrow -2\Gamma_{\rm cusp} \ln e^{\gamma_E}N-(\gamma_s+\gamma_V)+{\cal O}\left(\frac{1}{N}\right) \ .
\end{align}
All these anomalous dimensions are free from renormalons at the level of bubble chain diagrams. 

\subsection{Borel transform and conservation of ``genuine'' IR renormalons in~\cite{Braun:2018brg} }
To check consistency, one can also perform the threshold expansion at the level of the Borel-transforms of the renormalized results and verify the threshold factorization. Here we only present the results in coordinate space. Up to terms which are entire functions, by performing the threshold expansion on the $\overline {MS}$ scheme Borel transform,  in terms of the natural Borel argument $u=\frac{\beta_0t}{2}$, the first two orders in the large $\lambda$ expansion reads
\begin{align}
B\bigg[{\cal H}_{1a}(z^2,\lambda,\alpha(\mu))\bigg](u) = B^h_{0}(u) +B_0^J(u)+ \frac{1}{\lambda}B_{1}(u)  + O\left( \frac{1}{\lambda^2} \right).
\end{align}
In this equation, the first term
\begin{align}
&B_{0}^h(u) = \frac{e^{-i \lambda } C_F}{4 \pi u^2 } \left( e^{\frac{5u}{3}} \bigg(\frac{ i |z|\mu }{2\lambda}\right)^{2u} \frac{2\pi u}{(u-1)\sin 2\pi u} + \left(\frac{8}{3}+2\ln \frac{ i |z|\mu }{2\lambda}\right)u+1\bigg) \ ,
\end{align}
is nothing but the Borel transform of the hard kernel in Eq.~(\ref{eq:BorelHLfull}), (\ref{eq:BorelHLexp}).  The second term
\begin{align}\label{eq:BorelJ}
&B_0^J(u)= \frac{e^{-i \lambda } C_F}{4 \pi u^2 } \bigg( e^{\frac{5u}{3}} \left(\frac{z^2 \mu ^2}{4 }\right)^u \frac{\Gamma (1-u) }{\Gamma (1+u)}-\left(\frac{5}{3}+\ln \frac{e^{2\gamma_E}z^2\mu^2}{4}\right)u-1\bigg) \ ,
\end{align}
is the Borel transform of the (bubble chain diagrams between $n^+$ and $n_z$) space-like threshold soft factor. The $u=\frac{1}{2}$ singularity only appears in the hard contribution. The first two terms represent the leading power contribution in the threshold expansion. On the other hand, the term
\begin{align}
&B_{1}(u) = i\frac{e^{-i \lambda } C_F}{2\pi  u} \left( e^{\frac{5u}{3}} \left(\frac{z ^2 \mu ^2}{4 }\right)^u \frac{ \Gamma (1-u)}{(1-2 u) \Gamma (u+1)} -1\right) \ ,
\end{align}
is the NLP soft contribution. They are clearly in one-to-one correspondence with the three terms in Eq.~(\ref{eq:thresholdlimitc}) and can also be directly obtained from Eq.~(\ref{eq:thresholdlimitc}) using standard methods~\cite{Palanques-Mestre:1983ogz,Beneke:1994sw,Beneke:1995pq,Scimemi:2016ffw}. Again, the linear renormalon at $u=\frac{1}{2}$ cancels between LP hard kernel and NLP soft factor.

It is important to notice that although the $u=\frac{1}{2}$ renormalon for the hard kernel generated during the threshold expansion cancel between leading and next to leading threshold powers, the renormalons for the LP threshold soft factor is not cancelled by anything and exactly corresponds to the ``genuine'' renormalons in~\cite{Braun:2018brg} projected to the leading threshold power. They can only be cancelled by UV renormalons for high-twist operators (true ``non-perturbative'' effect) in the $z\Lambda_{\rm QCD}$ expansion. To show this we need to add to twice of Eq.~(\ref{eq:BorelJ}) the contribution from the gauge-link self interactions (namely, Fig.~1d of~\cite{Braun:2018brg})
\begin{align}
B^{\rm self}(u)=-\frac{e^{-i \lambda } C_F}{2 \pi u} \bigg( e^{\frac{5u}{3}} \left(\frac{z^2 \mu ^2}{4 }\right)^u \frac{\Gamma (1-u) }{(2u-1)\Gamma (1+u)}+1\bigg) \ .
\end{align}
Notice the striking similarity of the Borel transform for the self-energy contribution and the NLP soft factor. Combining the above, one has the Borel transform for all the three bubble chain diagrams for the LP threshold soft factor Eq.~(\ref{eq:defJ})
\begin{align}
B^{J}_{\rm total}(u)=\frac{e^{-i\lambda }C_F}{2\pi u^2 }\bigg( e^{\frac{5u}{3}} \left(\frac{z^2 \mu ^2}{4 }\right)^u\frac{(u-1) \Gamma (1-u)}{(2 u-1) \Gamma (u+1)}-\left(\frac{8}{3}+\ln \frac{e^{2\gamma_E}z^2\mu^2}{4}\right)u-1\bigg) \ .
\end{align}
Clearly, the above contains IR renormalons at $u=k$ for $k \ge 2$, with residues read
\begin{align}
{\rm Res}B^{J}_{\rm total}(u=k)=\frac{e^{-i\lambda }C_F}{2\pi } e^{\frac{5k}{3}} \left(\frac{z^2 \mu ^2}{4 }\right)^k \frac{(-1)^k (k-1)}{k^3 (2 k-1) \Gamma (k)^2} \ ,
\end{align}
which exactly corresponds to the $\delta(1-\alpha)$ term in Eq.~(49) of~\cite{Braun:2018brg} (the factor $4\pi$ difference is due to the fact that in~\cite{Braun:2018brg} the Borel transform is performed with respect to $\frac{\alpha_s}{4\pi}$). To conclude, the LP threshold soft factor Eq.~(\ref{eq:defJ}) correctly captures all the {\it genuine} IR renormalons for the pPDF OPE coefficients projected to the leading threshold power. There is essentially no enhancement of ``true'' non-perturbative effect due to scales softer than $\frac{1}{|z|}$ in the threshold limit.

\subsection{The NLP threshold soft factor and its UV renormalon} \label{sec: UVrenor}
In this subsection we address the nature of the NLP contribution Eq.~(\ref{eq:thresholdNLPcoor}). We show that it can be generated from a NLP threshold soft factor at the level of bubble chain diagrams. Moreover, we show that its  $s=\frac{1}{2}$ pole is UV.  

In fact, the NLP soft factor one needs simply reads (the NLP contribution for Fig.~1b of~\cite{Braun:2018brg} is identical. Fig.~1c starts contributing only at NNLP.)
\begin{align}\label{eq:NLPjet}
J_{q}(z,\alpha(\mu))=\frac{i}{2}\big\langle \Omega \big|\infty n^++zn_z, zn_z][zn_z,0]\int_{-\infty}^{0} dx^+[0,x^+]D_\perp^2[x^+,-\infty]\big|\Omega \big\rangle \ .
\end{align}
Here, $n^+=\frac{1}{\sqrt{2}}(1,0,0,1)$ is the light-front ``plus'' direction vector, and  $[y^+,x^+]$ denotes a light-like Wilson line in the $n^+$ direction from $x^+n^+$ to $y^+n^+$. 
This can be explained as the sub-eikonal insertion at the collinear quark line (explaining the subscript $q$ in the definition)
\begin{align}
\frac{i(p^++k^+)}{(p+k)^2+i0}=\frac{i}{2(k^-+i0)}+\frac{i}{p^+}\frac{k_\perp^2}{4(k^-)^2}+{\cal O}\left(\frac{1}{p^+}\right)^2 \ .
\end{align}
One can show that the bubble chain for $\frac{e^{-i\lambda}}{p^+}J_q(z,\alpha(\mu))$ reproduces the NLP threshold power contribution Eq.~(\ref{eq:thresholdNLPcoor}).  More precisely,  for $z>0$ one has
\begin{align}
\frac{e^{-i\lambda}}{p_+}J_{q,n}=g^2C_F\alpha^n\bigg(\frac{\beta_0}{2}\bigg)^n\mu_0^{2\epsilon}\mu^{2n\epsilon}e^{n\epsilon \gamma_E}\frac{1}{\epsilon^n}f(\epsilon)^n J_n \ ,
\end{align}
with 
\begin{align}
&J_n=-ie^{-i\lambda}\frac{i^{n\epsilon}n^+\cdot n_z}{2p^+\Gamma(1+s-\epsilon)}\int_{0}^{\infty}\alpha^{n\epsilon}d\alpha\int_{0}^{z}d\lambda_1\int_{0}^{\infty}\lambda_2d\lambda_2\int \frac{d^Dk}{(2\pi)^D}\vec{k}_\perp^2 \nonumber \\ 
& \times \exp \bigg[i\alpha k^2-ik\cdot(\lambda_1 n_z+\lambda_2n^+)\bigg] \ .
\end{align}
Now, after integration one has
\begin{align}
J_n=-\frac{ie^{-i\lambda}}{zp^z}\frac{2}{(4\pi)^{\frac{D}{2}}}\left(\frac{z^2}{4}\right)^s\frac{(\epsilon-1)\Gamma (-s)}{(2 s-1) \Gamma (-\epsilon+s+1)} \ ,
\end{align}
which is exactly the NLP contribution in Eq.~(\ref{eq:thresholdNLPcoor}) obtained from direct expansion. For $z<0$ one can see it reproduces Eq.~(\ref{eq:thresholdNLPcoor}) as well. 

Here we show that the linear $s=\frac{1}{2}$ pole for the $J_q$ above is UV in nature. For this purpose one can change the $\int_{0}^{\infty} \lambda_2 d\lambda_2$ into $\int_{a}^{\infty} \lambda_2 d\lambda_2$ with $a>0$ playing the role of a UV cutoff. This only changes the short distance singularities of the integral. Then, the full result reads
\begin{align}
e^{i\lambda}J_n(z,a)=&\frac{i}{(4\pi)^{\frac{D}{2}}p^z}\frac{(\epsilon-1) 2^{\frac{1}{2}-\frac{3 s}{2}} \Gamma (-s) (a z)^{s-1} }{s \Gamma (-\epsilon+s+1)} \nonumber \\ 
& \times \left(\sqrt{2} a s \, _2F_1\left(1-s,s-1;s;-\frac{z}{\sqrt{2} a}\right)-z \, _2F_1\left(1-s,s;s+1;-\frac{z}{\sqrt{2} a}\right)\right) \ .
\end{align}
As expected, the singularity at $s=\frac{1}{2}$ simply vanishes. Now, after power-expansion in $a$, one has
\begin{align}
e^{i\lambda}J_n(z,a)=&\frac{i}{(4\pi)^{\frac{D}{2}}p^z} \bigg(\frac{\sqrt{\pi } (\epsilon-1) 2^{\frac{3}{2}-3 s} a^{2 s-1}  \Gamma \left(\frac{1}{2}-s\right)}{\sin \pi s\Gamma (-\epsilon+s+1)}\bigg) %{\color{blue}\frac{1}{2(1-s)}} 
\nonumber \\ 
&-\frac{i}{zp^z}\frac{2}{(4\pi)^{\frac{D}{2}}}\left(\frac{z^2}{4}\right)^s\frac{(\epsilon-1)\Gamma (-s)}{(2 s-1) \Gamma (-\epsilon+s+1)}+{\cal O}(az) \ .
\end{align}
Clearly, the second term simply corresponds to the finite result evaluated at $a=0$ and the first term corresponds to the ``power UV-divergence'' due to the fact that the operator dimension for $J_q$ equals to $1$. The pole at $s=\frac{1}{2}$ cancel between the ``UV divergent'' term in the first line and the ``finite term'' in the second line, implying that the $s=\frac{1}{2}$ renormalon of the NLP soft factor is actually a {\it UV renormalon}. More precisely, this corresponds to the ambiguity of separating leading (UV divergence) and next to leading power (finite renormalized term) in $az$.   Thus, the renormalon cancellation pattern in the coordinate space threshold expansion is fully in agreement with the general pattern of renormalon cancellation between LP and NLP for controlled asymptotic expansions in UV limit.  

\subsection{Exponentially small terms and resurgence of the full distribution from threshold asymptotics}\label{sec:resurgent}
In fact, the full asymptotic expansion for Eq.~(\ref{eq:bubblecoord}) in the threshold limit reads
\begin{align}\label{eq:fullasympthre}
&e^{-\lambda}\mathfrak{h}_n(z^2,i\lambda)=\frac{ 2}{(4\pi)^{\frac{D}{2}}}\left(\frac{|z|}{2}\right)^{2s}\bigg(\frac{\pi}{\lambda^{2s}}\frac{(1-\epsilon)\Gamma(-s)}{\sin 2\pi s \Gamma(2-s-\epsilon)} \nonumber \\ 
&-\frac{\Gamma (-s)}{2s \Gamma (-\epsilon+s+1)}+\sum_{k=1}^{\infty}\frac{(-1)^k(\epsilon-1)\Gamma(-s)}{(2s-k)\Gamma(s-\epsilon+2-k)\lambda^{k}}\bigg) \nonumber \\ 
&+\frac{ 2e^{-\lambda}}{(4\pi)^{\frac{D}{2}}}\left(\frac{|z|}{2}\right)^{2s}(-\lambda)^{\epsilon-s}\bigg(\sum_{k=1}^{\infty}\frac{(1-\epsilon
)\Gamma(\epsilon+s-1)\Gamma(-s)}{\Gamma(\epsilon+s-k)\lambda^{k+1}}\bigg) \ ,
\end{align}
where in the last line we included all the exponentially small contributions related to $\alpha \rightarrow 0^+$ asymptotics.  Clearly, the threshold expansion in coordinate space projected only to algebraic terms is not convergent, but we will show that the ``Borel ambiguity'' for the $\frac{1}{\lambda}$ expansion of these terms cancels the branch ambiguities of $(-\lambda)^{\epsilon-s}$ for the exponentially small term. Notice the ``threshold scale'' for the exponentially small term is $\frac{\sqrt{\lambda}}{|z|}$. 

Here we should stress that the overall coefficient $(-\lambda)^{\epsilon-s}$ in the last line is in fact at the branch cut for $\lambda>0$ and requires a scheme to be regarded as an analytic function in the ${\rm Re}(\lambda)>0$ plane. This is due to the fact that only for ${\rm Re}(\lambda)<0$, this term is the leading large $\lambda$ asymptotics and in the ${\rm Re}(\lambda)>0$ case it can only be determined when the leading terms (threshold terms) are determined. Indeed, the infinite sum in the second line is not Borel summable and requires a scheme to be well-defined. The scheme choice for the Borel sum then transmutes to the branch choice for the $(-\lambda)^{\epsilon-s}$ term to make sure the total result is analytic in $\lambda$ in the whole complex plane, or equivalently, to make sure the $\alpha<0$ part is completely cancelled between the threshold and small $\alpha$ contributions. On the other hand, since in this case the Fourier transforms for the exponentially small contributions are only supported in $\alpha<0$, one can simply take the threshold contributions and truncate the Fourier transforms to $0<\alpha<1$ to obtain the full distribution in the $\alpha$ space. Notice in this region the Fourier transform for the infinite sum in the second line is free from ambiguities since Fourier transform introduces additional $\frac{1}{(k-1)!}$ and the convergence radius for the Fourier transform is equal to the convergence radius for the Borel sum for the $\frac{1}{\lambda}$ expansion. In this way only the threshold asymptotics in $\lambda$ space is needed to recover the full distribution due to the correlation between the algebraic and exponentially small terms, demonstrating the principle of ``resurgence''. 

To check the above statements,  it is convenient to Fourier transform back to the $\alpha$ space. Then the threshold terms read (they are supported in $\alpha \le 1$)
\begin{align}\label{eq:alphaspace}
& \mathfrak{h}_n^{\rm thr}(z^2,\alpha)=\frac{2}{(4\pi)^{\frac{D}{2}}}\left(\frac{|z|}{2}\right)^{2s}\bigg(\frac{(1-\epsilon)\Gamma(-s)\Gamma(1-2s)}{\Gamma(2-s-\epsilon)} \left(\left [\bar \alpha^{-1+2s} \right]^{[0,1]}_{+,1} + \frac{\delta(\bar \alpha)}{2s}\right) \nonumber \\ 
&-\frac{\Gamma (-s)}{2s \Gamma (-\epsilon+s+1)}\delta(\bar \alpha)+\frac{(1-\epsilon) \, _2F_1(1-2 s,\epsilon-s;2-2 s;\bar \alpha)}{(2 s-1) \Gamma (-\epsilon+s+1)} \Gamma (-s) \bigg)
\end{align}
Now, the contribution for the exponentially small contributions reads (it is supported only in $\alpha<0$)
\begin{align} \label{eq:smallxpart}
& \mathfrak{h}_n^{\rm sm}(z^2,\alpha)=\frac{ 2}{(4\pi)^{\frac{D}{2}}}\left(\frac{|z|}{2}\right)^{2s}\nonumber \\ 
& \times (\epsilon-1) \alpha  (-\alpha)^{s-\epsilon}(-1)^{s-\epsilon} \, _2\tilde{F}_1(1,-\epsilon-s+2;-\epsilon+s+2;\alpha)\Gamma (-s) \ .
\end{align}
To show that it compensates the threshold contributions for $\alpha<0$, it is convenient to use the connecting formula for hypergeometric functions that changes $1-\alpha$ to $\alpha$
\begin{align}
\frac{\sin \pi(c-a-b)}{\pi}\, _2\tilde F_1(a,b,c;1-\alpha)=&\frac{1}{\Gamma(c-a)\Gamma(c-b)}\, _2\tilde F_1(a,b,a+b-c+1;\alpha)\nonumber \\ 
&-\alpha^{c-a-b}\frac{\, _2\tilde F_1(c-a,c-b,c-a-b+1;\alpha)}{\Gamma(a)\Gamma(b)} \ ,
\end{align}
where $a=1-2s$, $b=\epsilon-s$, $c=2-2s$. Now, for $\alpha<0$ the original $\, _2F_1$ at the left-hand side of the equation has a branch cut, reflected in the $\alpha^{c-a-b}$ term. One can see from the above that if the branch for this $\alpha^{c-a-b}$ is chosen in agreement with the $(-1)^{s-\epsilon}$ in Eq.~(\ref{eq:smallxpart}), then the $\alpha<0$ part exactly cancels between Eq.~(\ref{eq:alphaspace}) and Eq.~(\ref{eq:smallxpart}). As a result, the full $\alpha$ space distribution is given by Eq.~(\ref{eq:alphaspace}) in $0<\alpha \le 1$. Notice that in the region $0<\alpha<1$, after applying the connecting formula,  Eq.~(\ref{eq:alphaspace}) is directly equal to Eq.~(A6) of~\cite{Braun:2018brg}. To show this, first converting the $\, _2F_1$ to the ``plus-distribution''
\begin{align}
\,_2F_1(1-2s,\epsilon-s,2-2s;\bar \alpha) =&\,_2F_1(1-2s,\epsilon-s,2-2s;\bar  \alpha)_{+,1}^{[0,1]}\nonumber \\ 
&-\frac{\Gamma (2-2 s) \left(\frac{2 s \Gamma (-\epsilon+s+1)}{\Gamma (-\epsilon-s+2)}-\frac{1}{(\epsilon-s-1) \Gamma (-2 s)}\right)}{4 s^2} \delta(\bar \alpha) \ .
\end{align}
Then combing all $\delta(\bar \alpha)$ terms one has in coordinate space
\begin{align}
&\mathfrak{h}_n(z^2,\lambda) = \int_{0}^{1} d\alpha \, e^{-i \alpha \lambda} \mathfrak{h}_n^{\rm thr}(z^2,\alpha) \nonumber\\
&= \int_{0}^{1} d\alpha \, e^{-i \alpha \lambda} \frac{2}{(4\pi)^{\frac{D}{2}}}\left(\frac{|z|}{2}\right)^{2s} \frac{\Gamma (-s)}{\Gamma (s-\epsilon +2)} \bigg[ \frac{(1-\epsilon ) \Gamma (1-2 s) \Gamma (s-\epsilon +2)}{\Gamma (-s-\epsilon +2)} [\bar \alpha^{-1+2s}]^{[0,1]}_{+,1} \nonumber\\ 
&\quad \quad +\frac{(\epsilon -1) (s-\epsilon +1) \, _2F_1(1-2 s,\epsilon -s;2-2 s; \bar \alpha )_{+,1}^{[0,1]}}{1-2 s} -\frac{1}{2} \delta (\bar \alpha ) \bigg]  \ . 
\end{align}
Now, using the connecting formula for the hypergeometric function, the $\bar \alpha^{-1+2s}$ term is cancelled, left finally with 
\begin{align}
&\mathfrak{h}_n(z^2,\lambda)=\frac{2}{(4\pi)^{\frac{D}{2}}}\left(\frac{|z|}{2}\right)^{2s} \frac{\Gamma (-s)}{\Gamma (s-\epsilon +2)} 
\nonumber \\ 
&\times \bigg[(1-\epsilon) \int_{0}^{1} d\alpha \,(e^{-i \alpha \lambda} - e^{-i \lambda})  \alpha ^{s-\epsilon +1} \, _2F_1(1,-s-\epsilon +2;s-\epsilon+2;\alpha ) - \, \frac{e^{-i \lambda}}{2} \bigg] \ ,
\end{align}
which is identical to Eq.~(A6) of~\cite{Braun:2018brg}.  In fact, the above procedure, when reverted, can also be used to derive the threshold limit. Nevertheless, the fact that we started from Eq.~(\ref{eq:bubblecoord}) and recovered the full distribution from threshold asymptotics provides a strong consistency check of the whole procedure as well. 

\subsection{A conjecture on threshold limit of hadronic quark-bilinear matrix element}

Notice that although up to now, we are discussing the threshold-limit of leading twist coefficient function for quark-bilinears, the results allow extension to threshold limit of the hadronic matrix element~\cite{Braun:2018brg} \begin{align}
\langle N(p)|\bar \psi(vz)\slashed{v}[vz,0]\psi(0)|N(p)\rangle=2v\cdot p {\cal F}(-v^2z^2,\lambda=-v\cdot p z) \ .
\end{align}
For simplicity we only consider the flavor non-singlet valence-quark induced hadronic matrix element. Clearly, for $z>0$ the above matrix element allows analytic continuation in $v$ with imaginary part in the negative forward light-cone, in particular, with $v=-i(1,0,0,0)$ in the Euclidean time direction. Compared with $v=n_z$, this does not change $-v^2=1$, but changes $\lambda \rightarrow i\lambda$, which is exactly the positive imaginary axis along which threshold limit can be extracted. Furthermore, using $x\alpha=1+(\alpha-1)+(x-1)+(1-\alpha)(1-x)$, one can see that in the large $\lambda \rightarrow +\infty$ limit, at leading twist, one has for the 
\begin{align}
&{\cal F}(z^2,i\lambda) \rightarrow e^{\lambda}H_{\rm HL}^2\bigg(0,\alpha(z_\lambda)\bigg)J(0,\alpha(z))\exp \bigg(\hat O_H(\alpha(z_\lambda))+\hat O_J(\alpha(z))\bigg) \nonumber \\ 
& \times \exp \bigg(2\ln \lambda e^{\gamma_E}\bigg(f_{\Gamma}(\alpha(\mu))-\frac{\gamma_0}{\beta_0}\ln \alpha(\mu)\bigg)+f_{s+V+2F}(\alpha(\mu))-\frac{\gamma^0_{s+V+2F}}
{\beta_0}\ln \alpha(\mu)\bigg) \nonumber \\ 
& \times {\cal Q}_{\rm PDF}(i\lambda,\alpha(\mu))\bigg(1+{\cal O} \left(\frac{1}{\lambda}\right)\bigg) \ ,
\end{align}
where ${\cal Q}_{\rm PDF}(i\lambda,\alpha(\mu))$ is the threshold limit of the quark PDF (without the $e^{\lambda}$ factor) and $z_\lambda=\frac{z}{\lambda}=\frac{1}{p^0}$ is the hard scale. Resummation factors are defined in Eq.~(\ref{eq:thresholdcon}), Eq.~(\ref{eq:resummationfactor}). In fact, due to the fact that the $\mu$ dependency cancels between the PDF and the $\alpha(\mu)$ part of the resummation factor, one can combine them to form the $\mu$-independent version of quark PDF, in term of which the leading twist threshold limit takes the form
\begin{align}
&{\cal F}(z^2,i\lambda) \rightarrow e^{\lambda}H_{\rm HL}^2\bigg(0,\alpha(z_\lambda)\bigg)J(0,\alpha(z))\exp \bigg(\hat O_H(\alpha(z_\lambda))+\hat O_J(\alpha(z))\bigg) \nonumber \\ 
& \times  {\cal Q}_{\rm PDF}(i\lambda)\bigg(1+{\cal O} \left(\frac{1}{\lambda}\right)\bigg)\times \exp \bigg(f_{2F}(\alpha(\mu))-\frac{\gamma^0_{2F}}
{\beta_0}\ln \alpha(\mu)\bigg) \ .
\end{align}
It is clear now that all the remaining $\mu$ dependency is due to the UV anomalous dimension of the quark-bilinear itself. 

Now, one may question if the leading twist always dominate the large $\lambda$ limit. In our opinion, we believe that the leading twist dominance holds to $\lambda \rightarrow +\infty$ in the sense of Eq.~(\ref{eq:thresholdhadron}) to be introduced. The reason is that, in the large $\lambda$ limit, due to the exponential suppression caused by the Euclidean time evolution, the large energy $\frac{\lambda}{z}$ of the incoming and out going hadron must flow out and in at the quark-line vertices, leaving the intermediate state (namely, the ``cut'') light. In this way, the large $\lambda$ limit is dominated by threshold contributions,  for which the two vertices are always in Sudakov-like hard kernels with at least one incoming collinear parton and one external soft gauge-link. Clearly, increasing the number of incoming collinear partons (except longitudinal gluons which factorize to eikonal lines in the PDF) and external soft partons to the hard kernel always introduce additional power suppressions. The Sudakov hard kernel with one incoming collinear valence quark and one external soft gauge-link is the hard kernel at leading power, on top of which collinear contributions factorize into the threshold limit of twist-two PDF (this explains the name of ``twist-two dominance'') and soft exchanges factorize into the threshold soft factor. Moreover, this picture for the dominant contribution in the large $\lambda $ limit should hold for moderate  $z$ as well, due to the fact that the threshold soft factor, defined as a Wilson-loop average, allows non-perturbative generalization. 

As such, we conjecture the following $\lambda \rightarrow +\infty$ factorization for the hadronic quark-bilinear matrix element from small to moderate $z$:
\begin{align}\label{eq:thresholdhadron}
&{\cal F}(z^2,i\lambda) \rightarrow e^{\lambda}H_{\rm HL}^2\bigg(0,\alpha(z_\lambda)\bigg)\exp \bigg(\hat O_H(\alpha(z_\lambda))\bigg)J(z\Lambda_{\rm QCD}) \nonumber \\ 
& \times  {\cal Q}_{\rm PDF}(i\lambda)\bigg(1+{\cal O} \left(\frac{1}{\lambda}\right)\bigg)\times \exp \bigg(f_{2F}(\alpha(\mu))-\frac{\gamma^0_{2F}}
{\beta_0}\ln \alpha(\mu)\bigg) \ .
\end{align}
Here $J(z\Lambda_{\rm QCD})$ is the appropriate non-perturbative version of the threshold-soft factor Eq.~(\ref{eq:defJ}) with $\mu$ dependency remove by evolution factors. The small $z$ expansion of the threshold soft factor is exactly 
\begin{align}
J(z\Lambda_{\rm QCD}) \rightarrow J(0,\alpha(z))\exp \bigg(\hat O_J(\alpha(z))\bigg)\bigg(1+{\cal O}(z^2\Lambda_{\rm QCD}^2)\bigg) \ .
\end{align}
Clearly, the fully factorized form of the threshold limit provides considerable simplification compared to the leading twist factorization formula at generic $\lambda$. Provided that ${\cal Q}_{\rm PDF}(i\lambda)$ does not decay very fast in the $\lambda \rightarrow +\infty$ limit, in principle it allows the lattice verification of the celebrated ``marginally-relevant'' Sudakov-like asymptotics Eq.~(\ref{eq:resummationfactor}) which dates back to late 1970s~\cite{Mueller:1979ih,Sen:1981sd}.

\section{Renormalon cancellation and power correction to TMD limit of quasi-LFWF amplitude}\label{sec:LFWF}
In this section, we consider the role played by the linear renormalon of the hard kernel in the TMD expansion of quasi-LFWF amplitude. We show that this linear renormalon cancels with UV renormalon of NLP soft contribution. We also show that in this case the TMD expansion fails to converge to the full result due to exponentially small contributions. 

As a reminder, the quasi-LFWF is defined as the Fourier transform of the following correlator~\cite{Ji:2020ect,Ji:2021znw}
\begin{align}\label{eq:quasiWF1}
&\widetilde \Psi^{\pm }(\lambda,b_\perp,\mu,4P_z^2)=\!\lim_{L\to\infty}\frac{\langle 0|\bar \psi\big(zn_z\!+\!\vec{b}_\perp\big)\gamma^z{\cal W}_{\mp z}(zn_z\!+\!\vec{b}_\perp;L)\psi\big(0\big)|P\rangle}{\sqrt{Z_E(2L,b_\perp,\mu)}} \ ,
\end{align}
where $\lambda=zP^z$ is the natural scaleless Lorentz invariant length conjugating to the quark's momentum fraction $x$. The staple-shaped gauge-link ${\cal W}_z$ is defined as 
\begin{align}\label{eq:staplez}
{\cal W}_{\pm z}(\xi;L)= [\xi,\mp Ln_z+\vec{b}_\perp][\mp Ln_z+\vec{b}_\perp,\mp Ln_z][\mp Ln_z,0] \ .
\end{align}
On the other hand, $Z_E$ is the square root of the vacuum expectation value of a flat rectangular Euclidean Wilson-loop
\begin{align}\label{eq:Z_E}
Z_E(2L,b_\perp,\mu)=\frac{1}{N_c}{\rm Tr}\langle 0|W_{\perp}{\cal W}_z(\vec{b}_\perp;2L)|0\rangle \ ,
\end{align}
that subtracts out the gauge-link self-interactions in the large $L$ limit. The $Z_E$ itself has nothing to do with the factorization. 

As shown in Ref.~\cite{Ji:2021znw}, the quasi-LFWF amplitude has the following leading-power factorization in the double-logarithmic evolution limit $P_z^2 \rightarrow \infty$ with fixed $x$ and $b_\perp$:
\begin{align}\label{eq:quai_fac_1}
\tilde \Psi^{\pm}_{\bar q q}(x,b_\perp,\mu,\zeta_z)S_{r}^{\frac{1}{2}}(b_\perp,\mu) =H^\pm\left(\frac{\zeta_z}{\mu^2},\frac{\bar \zeta_z}{\mu^2},\alpha(\mu)\right)e^{\frac{1}{2}\ln \frac{\mp \zeta_z+i0}{\zeta}K(b_\perp,\mu)}\Psi_{\bar qq}^\pm(x,b_\perp,\mu,\zeta) \ ,
\end{align}
where the hard kernel $H^{\pm}$ is nothing but a product of two heavy-light Sudakov form factors in Eq.~(\ref{eq:defHLsuda})
\begin{align}\label{eq:Hplusmins}
H^\pm\left(\frac{\zeta_z}{\mu^2},\frac{\bar \zeta_z}{\mu^2},\alpha(\mu)\right)=H_{\rm HL}\bigg(L_z,\mp,\alpha(\mu)\bigg)H_{\rm HL}\bigg(\bar L_z,\mp,\alpha(\mu)\bigg) \ .
\end{align}
The natural scales $\zeta_z=4x^2P_z^2$ and $\bar \zeta_z=4(1-x)^2P_z^2$ are provided  by the momenta $p=xP$ of the collinear quark and $\bar p=(1-x)P$ of anti-quark. 

In this section, we only consider the quasi-LFWF amplitude in PQCD for the external states with a pair of on-shell quark anti-quark (with momenta $p^2=\bar p^2=0$), which serves as the small $b$ OPE coefficients that match the mesonic quasi-LFWF to collinear distribution amplitudes (DA). We calculate the bubble chain diagrams for the quasi-LFWF amplitude that couples the external quark and anti-quark to the space-like gauge links, since at bubble chain level only these two diagrams know about the presence of multiples scales at LP. To avoid complications caused by the $\lambda$ (or momentum fractions), we simply calculate the $\lambda=0$ version of the quasi-LFWF amplitude. It is not hard to see that at the bubble-chain level, the factorization for $\lambda=0$ simply reduces to additions:
\begin{align}
\tilde \Psi^{\pm}(b_\perp,\mu,\zeta_z)=H^\pm\left(\frac{\zeta_z}{\mu^2},\frac{\bar \zeta_z}{\mu^2},\alpha(\mu)\right)+\frac{1}{2}\ln \frac{\mp \zeta_z+i0}{\zeta}K(b_\perp,\mu)+{\rm collinear+soft} \ .
\end{align}
Here $\lambda=0$ is omitted for simplicity. Below we calculate the bubble chain diagram for the $\lambda=0 $ quasi-LFWF $\tilde \Psi^+$ (the minus version simply differs in sign for the imaginary part) and demonstrate the renormalon cancellation between leading and next to leading power. 

\subsection{Bubble chain diagram for quasi-LFWF amplitude}
Here we consider the bubble chain diagram for the quasi-LFWF amplitude at $\lambda=0$. As explained before, we only consider the two bubble chain diagrams which couple the external quark/anti-quark line with the space-like gauge link.  For quasi-LFWF amplitude, two diagrams contribute equally with replacements $\zeta_z \rightarrow \bar \zeta_z$, so we only present the calculation details for one of them, the one which connects the collinear quark to the Wilson line. The diagram contains two scales, $\zeta_z=4p_z^2$ and $b_\perp$. Using the same notation as Eq.~(\ref{eq:fnnotation}), introducing
\begin{align}
\tilde \Psi^+=g^2C_F\alpha^n\bigg(\frac{\beta_0}{2}\bigg)^n\mu_0^{2\epsilon}\mu^{2n\epsilon}e^{n\epsilon \gamma_E}\frac{1}{\epsilon^n}f(\epsilon)^n \psi_n \ ,
\end{align}
one has the integral representation
\begin{align}
&\psi_n=-\frac{2i}{(4\pi)^{\frac{D}{2}}(4p_z^2)^{(n+1)\epsilon}\Gamma(1+n\epsilon)}\nonumber \\ 
&\times\int_{0}^{\infty}d\lambda \int_{0}^{\infty} d\rho \rho^{-1+(n+1)\epsilon}\int_0^1dx \bar x^{n\epsilon}e^{-\frac{\lambda^2}{\rho}-i\lambda x} (1-e^{-\frac{\zeta_zb_\perp^2}{4\rho}})\bigg(1-x+\frac{i\lambda}{\rho}\bigg) \ .
\end{align}
Now we consider the imaginary part (as one can check, only the imaginary part contributes to NLP at $\frac{1}{\sqrt{\zeta_z}b_\perp}$), for which the $\lambda$ integral can be evaluated exactly, leading to 
\begin{align}\label{eq:Imfn}
&{\rm Im}\psi_n\nonumber\\&
=-\frac{\sqrt{\pi}}{2(4\pi)^{\frac{D}{2}}(4p_z^2)^{s}\Gamma(1+s-\epsilon)}\int_{0}^{\infty} d\rho \rho^{-\frac{1}{2}+s}\int_0^1dx \bar x^{s-\epsilon}e^{-\frac{\rho x^2}{4}} (1-e^{-\frac{\zeta_zb_\perp^2}{4\rho}})\bigg(2-x\bigg) \ ,
\end{align}
where one have used $s=(n+1)\epsilon$. Now, the integral for $\rho$ can be further performed, leading to 
\begin{align}
&{\rm Im}\psi(\epsilon,s)=-\frac{\sqrt{\pi}}{2(4\pi)^{\frac{D}{2}}(4p_z^2)^{s}\Gamma(1+s-\epsilon)}\nonumber \\ 
&\times\int_{0}^1dx \bar x^{s-\epsilon}(2-x)\bigg(\left(\frac{2}{x}\right)^{1+2s}\Gamma\left(\frac{1}{2}+s\right)-2\left(\frac{\sqrt{\zeta_z}b_\perp}{x}\right)^{\frac{1}{2}+s}K_{\frac{1}{2}+s}\left(\frac{\sqrt{\zeta_z}b_\perp x}{2}\right)\bigg) \ .
\end{align}
It is easy to see that before performing the power-expansion, the above function is analytic at $s=\frac{1}{2}$ and $\epsilon=0$, therefore not contributing to linear renormalon. It still contributes to quadratic renormalon at $s=1$. This implies that in the deeply UV region $\zeta_z\sim\frac{1}{b_\perp^2}$, the OPE coefficient that matches the quasi-LFWF to the collinear DA has no linear renormalon.  The linear renormalon is purely a result of the $\zeta_z\gg \frac{1}{b_\perp}$ expansion in the double-logarithmic limit.
\subsection{Power expansion and renormalon cancellation}
One expects that the linear renormalon in the HL Sudakov cancels with linear renormalon in ``high-twist'' TMDWF amplitudes or soft factors. Indeed, if one perform the $b_\perp\ll \frac{1}{\Lambda_{QCD}}$ expansion, then the renormalon ambiguity of twist-three distribution will be of the structure $\frac{1}{b_\perp p_z} \times b_\perp \Lambda_{QCD}$, where $\frac{1}{b_\perp p_z}$ is the natural power for all the perturbative twist-three contributions, and $b_\perp\Lambda_{\rm QCD}$ is the linear renomalon in the twsit-three TMD distributions. This cancellation can actually be demonstrated. Indeed, for this purpose, it is convenient to perform Mellin transform on $q=\zeta_ zb_\perp^2$ in Eq.~(\ref{eq:Imfn}).  Using
\begin{align}
1-e^{-\frac{q}{4\rho}}=\rightarrow -(4\rho)^t\frac{\Gamma(1+t)}{t} \ ,  \ -1<{\rm Re }(t)<0 \ .
\end{align}
the Mellin transform reads
\begin{align}
 {\rm Im} \psi_n(t)=-\frac{\sqrt{\pi}}{2(4\pi)^{\frac{D}{2}}(4p_z^2)^{s}}
 \frac{2^{2 s+4 t+2}(\epsilon+t-1)  \Gamma (t) \Gamma \left(s+t+\frac{1}{2}\right) \Gamma (-2s -2t)}{\Gamma (-\epsilon-s-2 t+2)} \ ,
\end{align}
with the strip of convergence
\begin{align}
{\rm max}\{-1,-\frac{1}{2}-s \}<{\rm Re}(t)<-s \ ,
\end{align}
and the condition $\epsilon>-\frac{1}{n}$. It is easy to see now, that the heavy-light Sudakov contribution at leading power Eq.~(\ref{eq:HLfn}) exactly corresponds to the pole at $t=0$, while the IR contribution at leading power corresponds to $t=-s$ :
\begin{align}
 &{\rm Im} \psi_n(t \rightarrow 0)=-\frac{\sqrt{\pi}}{2(4\pi)^{\frac{D}{2}}\left(4 p_z^2\right){}^{s}}\frac{ 2^{2s+2}(\epsilon -1) \Gamma (-2 s) \Gamma \left(s+\frac{1}{2}\right)}{\Gamma (-s-\epsilon +2)t}  \ , \\
 &{\rm Im} \psi_n(t \rightarrow -s)=\frac{\sqrt{\pi}}{2(4\pi)^{\frac{D}{2}}(4p_z^2)^{s}} \frac{\sqrt{\pi } 2^{1-2 s} (-s+\epsilon -1) \Gamma (-s)}{\Gamma (s-\epsilon +2)(t+s)} \ .
\end{align}
The NL power correction corresponds to the pole at $t=-s+\frac{1}{2}$, with residue
\begin{align}
 {\rm Im} \psi_n\left(t\rightarrow -s+\frac{1}{2}\right)=\frac{\sqrt{\pi}}{2(4\pi)^{\frac{D}{2}}(4p_z^2)^{s}\Gamma (1+s-\epsilon)}\frac{2^{2-2 s} (-2\epsilon+2 s+1)}{t+s-\frac{1}{2}}\Gamma \left(\frac{1}{2}-s\right) \ .
\end{align}
Given this, contributions at various powers can be obtained now, by shifting the vertical line for Mellin inverse transform 
\begin{align}
{\rm Im}\psi_n=\frac{1}{2\pi i}\int_{\gamma-i\infty}^{\gamma+i\infty}{\rm Im}\psi_n(t)(4p_z^2b_\perp^2)^{-t}dt \ ,  \ {\rm max}\big\{-1,-\frac{1}{2}-s \big\}<\gamma<-s  \ ,
\end{align}
to the right of poles and pick up the residue at the pole. More precisely, the leading power contribution reads
\begin{align}
{\rm Im }\psi^{(0)}(\epsilon,s)=&\frac{\sqrt{\pi}}{2(4\pi)^{\frac{D}{2}}\left(4 p_z^2\right){}^{s}}\frac{ 2^{2s+2}(\epsilon -1) \Gamma (-2 s) \Gamma \left(s+\frac{1}{2}\right)}{\Gamma (-s-\epsilon +2)}\nonumber \\ 
&-\frac{\pi}{2(4\pi)^{\frac{D}{2}}} \frac{\sqrt{\pi } 2^{1-2 s} (-s+\epsilon -1) \Gamma (-s)}{\Gamma (s-\epsilon +2)}(b_\perp^2)^{s} \ .
\end{align}
The first line corresponds to the hard contribution while the second line corresponds to the soft contribution. The NLP contribution reads
\begin{align}\label{eq:WFf1}
{\rm Im} \psi^{(1)}(\epsilon,s)=-\frac{1}{(4p_z^2)^s}\frac{2^{2-2s}\sqrt{\pi}(-2\epsilon+2s+1)}{2(4\pi)^{\frac{D}{2}} \Gamma (1+s-\epsilon)}\Gamma\left(\frac{1}{2}-s\right)(4p_z^2b_\perp^2)^{s-\frac{1}{2}} \ .
\end{align}
It has the desired linear renormalon at $s=\frac{1}{2}$ and one can show that it precisely cancels the linear renormalon in the leading power hard contribution. Indeed, at leading power one has for the quasi-LFWF
\begin{align}
{\rm Im} \tilde \Psi^{(0)}(\epsilon,s)&=C_F\alpha^{n+1}\left(\frac{\beta_0}{2}\right)^n\frac{G^{(0)}(\epsilon,s)}{(n+1)\epsilon^{n+1}} \ , \\ 
 G^{(0)}(\epsilon,s)&=\frac{1}{2\sqrt{\pi}}\frac{ 2^{2s}(\epsilon -1) s\Gamma (-2 s) \Gamma \left(s+\frac{1}{2}\right)}{\Gamma (-s-\epsilon +2)}e^{\gamma_Es}f(\epsilon)^{\frac{s}{\epsilon}-1}\left(\frac{\mu^2}{4p_z^2}\right)^s \nonumber \\
&-\frac{(-s+\epsilon -1) s\Gamma (-s)}{4\Gamma (s-\epsilon +2)}e^{\gamma_Es}f(\epsilon)^{\frac{s}{\epsilon}-1}\left(\frac{\mu^2b_\perp^2}{4}\right)^{s} \ .
\end{align}
At NLP one has
\begin{align}
{\rm Im} \tilde \Psi^{(1)}(\epsilon,s)=\frac{1}{p_zb_\perp}C_F\alpha^{n+1}\left(\frac{\beta_0}{2}\right)^n\frac{G^{(1)}(\epsilon,s)} {(n+1)\epsilon^{n+1}} \ , \\ 
G^{(1)}(\epsilon,s)=-e^{\gamma_Es}\left(\frac{\mu^2b_\perp^2}{4}\right)^s f(\epsilon)^{\frac{s}{\epsilon}-1}\frac{(-2\epsilon+2 s+1)s \Gamma \left(\frac{1}{2}-s\right) }{4 \sqrt{\pi } \Gamma (-\epsilon+s+1)} \ .
\end{align}
Furthermore, one can check that the real part has no contribution at power $\frac{1}{p_zb_\perp}$.  This implies that the Borel transform of the LFWF at LP reads ($u=\frac{\beta_0 t}{2}$, $L_b=\ln \frac{\mu^2b_\perp^2}{4}$)
\begin{align}
{\rm Im }\tilde \Psi^{0}(u,b_\perp,\zeta_z)= \frac{1}{4}C_Fe^{\frac{5}{3}u}\bigg(\frac{ e^{L_bu} \Gamma (-u)}{\Gamma (u+1)}-\frac{e^{-L_zu} }{u(u-1)}\frac{1}{\cos \pi u}\bigg) \ ,
\end{align}
with residue at $u=\frac{1}{2}$ reads
\begin{align}
{\rm Im }\tilde \Psi^{0}(u,b_\perp,\zeta_z)=-\frac{\mu}{p_z}\frac{C_Fe^{\frac{5}{6}}}{2\pi \left(u-\frac{1}{2}\right)} \ .
\end{align}
This is the same as Eq.~(\ref{eq:Borel}) for $z<0$. On the other hand, the Borel transform at NLP reads 
\begin{align}\label{eq:NLPresults}
{\tilde \Psi}^{(1)}(u,b_\perp)=\frac{iC_F}{p_zb_\perp}\frac{G^1\left(0,u\right)}{u}=\frac{iC_F}{2p_zb_\perp}e^{\frac{5u}{3}}(\mu^2b_\perp^2)^u \frac{(2u+1) \sin (\pi  u) \Gamma (-2 u)}{\pi } \ ,
\end{align}
with the residue at $u=\frac{1}{2}$ reads
\begin{align}\label{eq:NLPresidue}
{ \tilde \Psi}^{(1)}(u)\sim \frac{i\mu}{p_z}\frac{C_Fe^{\frac{5}{6}}}{2\pi \left(u-\frac{1}{2}\right)} \ .
\end{align}
It cancels the linear renormalon at leading power. Moreover, the Borel transform is almost identical to the bubble chain diagram for the heavy-quark potential, except for the $2u+1$ term. 

Note that the above analytical results may improve the accuracy in extracting the CS kernel, especially in suppressing the imaginary part. See Appendix~\ref{sec:ImCSK} for more discussions. 

\subsection{NLP soft factor. Linear UV vs rapidity divergences}
Now the question arises, can one attribute the above NLP contribution to twist-three TMD distributions or soft factors?  In fact, the NLP contribution at the bubble chain level\footnote{As a reminder, in this section we only consider the bubble chain diagram for the quasi-LFWF at $\lambda=0$. An enumeration of all the NLP contributions to the quasi-LFWF is outside the scope of this paper, as most of them do not correlate with renormalons at LP. For quasi-TMDPDF, an enumeration has been given in Ref.~\cite{Rodini:2022wic}, for which the contribution of $S_q$ and $S_{\bar q}$ cancels and does not appear in the final result. } can be explained by the soft factor
\begin{align}\label{eq:NLPsoft}
S^{\pm}_{q}(b_\perp,L,\mu)=\frac{i}{2}\big\langle \Omega \big|[-\infty n^++b_\perp,b_\perp] {\cal W}_{\mp z}(\vec{b_\perp};L)\int_{-\infty}^{0} dx^+[0,x^+]D_\perp^2[x^+,-\infty]\big|\Omega \big\rangle \  ,
\end{align}
which is similar to Eq.~(\ref{eq:NLPjet}) for pPDF in the threshold limit. Given the above, it is not difficult to show that the bubble-chain diagram for $\frac{1}{p^+}S_q$ exactly reproduces the NLP contribution Eq.~(\ref{eq:NLPresults}). Another bubble chain diagram between the antiquark and space-like Wilson-line has a similar $S_{\bar q}$ soft factor which contributes the same as $S_{q}$ in case of quasi-LFWF but cancels with $S_q$ for quasi-TMDPDF. The only scale for $S_q$ is $b_\perp$, explaining its name as a ``soft factor''. More precisely, one has
\begin{align}
\frac{1}{p_+}S_{q,n}^-=ig^2C_F\alpha^n\bigg(\frac{\beta_0}{2}\bigg)^n\mu_0^{2\epsilon}\mu^{2n\epsilon}e^{n\epsilon \gamma_E}\frac{1}{\epsilon^n}f(\epsilon)^n s_{q,n}^- \ ,
\end{align}
with 
\begin{align}
&s_{q,n}^-=\frac{i^{n\epsilon}n^+\cdot n_z}{2p^+\Gamma(1+n\epsilon)}\eta^{\nu}\int_{0}^{\infty}\alpha^{n\epsilon}d\alpha\int_{0}^{\infty}d\lambda_1\int_{0}^{\infty}\lambda_2^{1+\nu }d\lambda_2\int \frac{d^Dk}{(2\pi)^D}\vec{k}_\perp^2 \nonumber \\ 
& \times \exp \bigg[i\alpha k^2+i\vec{k}\cdot \vec{b}_\perp-ik\cdot(\lambda_1 n_z+\lambda_2n^+)\bigg] \ .
\end{align}
In this above, $\nu$ is the rapidity regulator which should be sent to $0$ later, $\eta$ is a mass-scale similar to the $\delta^+$ in the delta regulator. Now, after Wick rotation and performing the integrals, one has
\begin{align}
s_{q,n}^-=-\frac{2^{2-2s-\frac{\nu}{2}}b_\perp^{2s-1+\nu}(2s-2\epsilon+1+\nu)\eta^{\nu}\Gamma(2+\nu)}{8p^z(4\pi)^{\frac{D}{2}}\Gamma(s-\epsilon+1)}\Gamma\left(-\frac{1}{2}-\frac{\nu}{2}\right)\Gamma\left(\frac{1}{2}-s-\frac{\nu}{2}\right) \ .
\end{align}
From this, it is clear that the $\nu \rightarrow 0$ limit is regular and will not generate any logarithmic rapidity divergence. Sending $\nu$ to zero, one has
\begin{align}
s_{q,n}^{\pm}=\mp \frac{\sqrt{\pi}(2s-2\epsilon+1)\Gamma\left(\frac{1}{2}-s\right)}{p^zb_\perp(4\pi)^{\frac{D}{2}}\Gamma(s-\epsilon+1)}\left(\frac{b_\perp^2}{4}\right)^s \ ,
\end{align}
which simply reproduces the results Eq.~(\ref{eq:WFf1}) before. 

A special feature of the above NLP soft factor is that in its definition one needs a ``high-dimensional Wilson-line cusp'' 
\begin{align}
\hat O=[n_z\infty,0]\int_{-\infty}^{0} dx^+[0,x^+]D_\perp^2[x^+,-\infty] \ ,
\end{align}
with dimension $1$, therefore in principle can contribute to power-like UV divergences. Moreover, the correlator Eq.~(\ref{eq:NLPsoft}) suffers from {\it linear rapidity divergence} which requires special treatment. Although in regulators without introducing new scales into Feynman integrals such as the {\it analytic regulator}~\cite{Becher:2010tm} adopted above, the linear rapidity divergence in Eq.~(\ref{eq:NLPsoft}) is simply removed when removing the regulator $\nu \rightarrow 0$ like the linear divergence in DR when $\epsilon \rightarrow 0$, the existence of such divergence strongly suggests that the object in definition still suffers rapidity regularization scheme dependency. So which one among these two linear divergences is related to the linear UV renormalon? The answer is, it is not the linear rapidity divergence, but the linear UV divergence that correlates with the linear renormalon. Indeed, one can introduce a displacement $a$ in $z$ direction in a way similar to the {\it exponential regulator}~\cite{Li:2016axz,Moult:2018jzp} to regulate the rapidity divergence as well. It is not hard to show that the bubble chain diagram after introducing this regulator still contains $s=\frac{1}{2}$ pole at $a^0$ that can not be canceled by ``linear rapidity divergences'' at $a^{-1}$. Only after introducing the cutoff regulator for the ``high-dimensional Wilson-line cusp'' in a way similar to that in Sec.~\ref{sec: UVrenor} the $s=\frac{1}{2}$ pole can be removed, demonstrating the UV nature of the renormalon. Thus, the linear rapidity divergence is not sufficient to generate linear renormalons, at least in the current case.

\subsection{Exponentially small contributions and divergence of TMD expansion}  In this example, one can make the interesting observation: for any $s$, the power expansion based on Mellin's transform
\begin{align}
 {\rm Im} \psi(s,t)=-\frac{\sqrt{\pi}}{2(4\pi)^{\frac{D}{2}}(4p_z^2)^{s}}
 \frac{2^{2 s+4 t+2}(\epsilon+t-1)  \Gamma (t) \Gamma \left(s+t+\frac{1}{2}\right) \Gamma (-2s -2t)}{\Gamma (-\epsilon-s-2 t+2)} \ ,
\end{align}
is not convergent. In fact, the power-expansion corresponds to poles at
\begin{align}
t_n=-s+\frac{n}{2} \ ,
\end{align}
with $n\ge 0$.  The non-convergence of the OPE can be read from the negative residue
\begin{align}
A_n= -\frac{\sqrt{\pi}}{2(4\pi)^{\frac{D}{2}}(4p_z^2)^{s}}\frac{(-1)^n 2^{2 n-2 s} \Gamma \left(\frac{n+1}{2}\right) (2 \epsilon+n-2 s-2) \Gamma \left(\frac{n}{2}-s\right)}{n!  \Gamma (-\epsilon-n+s+2)} \ ,
\end{align}
which diverges factorialy in the $n\rightarrow \infty$ limit. The above implies that the quasi-LFWF amplitude at power $(p^zb_\perp)^{-n}, n\ge 1$  has the Borel transform
\begin{align}
{\rm Im} \tilde \Psi^{(n)}(u,b_\perp)=\frac{C_F}{(p^zb_\perp)^n}(-1)^{n}\frac{(2u+2-n)\Gamma\left(\frac{n}{2}-u\right)}{8\Gamma \left(1+\frac{n}{2}\right)\Gamma(u+2-n)}e^{\frac{5}{3}u+L_b u} \ ,
\end{align}
which implies that the power expansion diverges for a generic $u>0$. For example, the derivative of the above at $u=0$ reads for $n \ge 2$
\begin{align}
{\rm Im} \frac{d}{du } \tilde \Psi^{(n)}(u,b_\perp)|_{u=0}=-\frac{C_F}{4(p^zb_\perp)^n}\frac{(n-2)\Gamma(n-1)}{n} \ ,
\end{align}
which growth factorially. The Borel transform of the above reads
\begin{align}
\sum_{n=2}^{\infty}{\rm Im} \frac{d}{du } \tilde \Psi^{(n)}(u,b_\perp)|_{u=0}\frac{t^n}{n!}=\frac{\left(-2 b_\perp p^z \text{Li}_2\left(\frac{t}{b_\perp p^z}\right)+(b_\perp p^z-t) \ln\left(1-\frac{t}{b_\perp p^z}\right)+3 t\right)}{4 b_\perp p^z} \ ,
\end{align}
which has a singularly at $t=b_\perp p^z$, suggesting naruall correlation with exponential small contributions at $e^{-b_\perp p^z}$.

In fact, the OPE non-convergence as well as existence of such an exponential small contribution can be shown even at one-loop level for $s=\epsilon$. In this case, the imaginary part of the LFWF is UV
finite and has the representation
\begin{align}\label{eq:Imf0}
&{\rm Im}\psi_0
=-\frac{\sqrt{\pi}}{32\pi^2}\int_{0}^{\infty} d\rho \rho^{-\frac{1}{2}}\int_0^1dx e^{-\frac{\rho x^2}{4}} (1-e^{-\frac{\zeta_zb_\perp^2}{4\rho}})\bigg(2-x\bigg)  \nonumber \\ 
&=-\frac{1}{8\pi}\int_{0}^1\frac{1}{x}\bigg(1-e^{-\frac{\sqrt{\zeta_z}b_\perp}{2}x}\bigg)+\frac{1}{16\pi}\bigg(1-\frac{2}{\sqrt{\zeta_z}b_\perp}\bigg)+\frac{1}{8\pi}\frac{1}{\sqrt{\zeta_z}b_\perp}e^{-\frac{\sqrt{\zeta_z}b_\perp}{2}}\ .
\end{align}
Now, after a few simplifications, the integral reads
\begin{align}
{\rm Im}\psi_0=&-\frac{1}{8\pi}\bigg(\gamma_E+\ln \frac{\sqrt{\zeta_z}b_\perp}{2}\bigg)+\frac{1}{16\pi}\bigg(1-\frac{2}{\sqrt{\zeta_z}b_\perp}\bigg) \nonumber \\ 
&-\frac{1}{8\pi}\int_{1}^{\infty}\frac{dx}{x}e^{-\frac{\sqrt{\zeta_z}b_\perp}{2}x}+\frac{1}{8\pi}\frac{1}{\sqrt{\zeta_z}b_\perp}e^{-\frac{\sqrt{\zeta_z}b_\perp}{2}} \ .
\end{align}
Clearly, the first line corresponds to the standard power expansion at leading power and NLP, while the second line is exponential-small contributions that can not be cast into the form of a power expansion. They are caused by ``tunneling'' of hard momenta 
between the two spatially separated hard kernels at $0$ and $\vec{b}_\perp$. On the other hand, when Fourier transformed into $k_\perp$ space,  such exponentially small contributions will result in algebraically power-suppressed terms starting from power $\frac{1}{p_z^2}$. This implies that the power expansion in $b_\perp $ space and $k_\perp$ space is not likely to be commuting starting at quadratic power. Moreover, compared to the case of 1D threshold-limit in Sec.~\ref{sec:threshold}, 
reconstruction of exponentially small terms from algebraic asymptotics for TMD expansion, due to the 2D nature, is less trivial even in the presence of support property (such as the Wightman-like current-current correlator for DY process, for which one has $|q_\perp| \le q^0$ due to spectral condition).  As such,  the $b_\perp$ space TMD expansion,  due to the clean separation between ``universal'' algebraic terms and ``non-universal'' exponentially small terms,  has the advantage in terms of distilling the massless scaling limit. On the other hand, the $k_\perp$ space TMD expansion condenses more information into power-log terms at small $k_\perp$ and has better phenomenological performance.

\section{Linear renormalon of quark wave function renormalization in Coulomb gauge.}\label{sec:coulomb}
Here we study the hard kernel for the recently proposed Coulomb-gauge approach to the lattice-TMD~\cite{Gao:2023lny,Zhao:2023ptv}. In this case, the gauge link is replaced by the following gauge transformation (called ``dressing factor'' in~\cite{Zhao:2023ptv} and references therein) which transforms a generic gauge to Coulomb gauge
\begin{align}\label{eq:CGDS}
U_{C}(x)=\exp \bigg(-\frac{ig}{\nabla^2}\nabla \cdot A(x)+{\cal O}(g^2)\bigg) \ .
\end{align}
Notice that in QED the ${\cal O}(g^2)$ corrections vanishes. The Coulomb gauge hard kernel is defined as the following object
\begin{align}
Z(p_0^2\mu^2,\alpha(\mu))u(p)=\langle 0|U_{C}(0)\psi(0)|p\rangle \ ,
\end{align}
where all the divergences are regulated in DR and subtracted {\it multiplicatively} in the $\overline{\rm MS}$ scheme. As for the heavy-light Sudakov hard kernel, the above Coulomb gauge hard kernel can be interpreted as a gauge-invariant definition of the {\it Coulomb gauge quark field renormalization}. 

We now calculate the bubble chain diagram of the above hard kernel. The diagram with $n$ bubble-chain insertions reads
\begin{align}
V_n^{C}=g^2C_F\alpha^n\bigg(\frac{\beta_0}{2}\bigg)^n\mu_0^{2\epsilon}\mu^{2n\epsilon}e^{n\epsilon \gamma_E}\frac{1}{\epsilon^n}f(\epsilon)^n h^C_n u(p) \ ,
\end{align}
where one has
\begin{align}
h^C_n u(p)=\int \frac{d^Dk}{(2\pi^D)}\frac{i}{(-k^2-i0)^{1+n\epsilon}}\frac{\slashed{p}-\slashed{k}}{(p-k)^2+i0}\frac{\slashed{\vec{k}}}{|\vec{k}|^2}u(p) \ .
\end{align}
Now, for $p=(p_0,0,0,p_0)$, using rotational symmetry and equation of motion, one can write 
\begin{align}
(\slashed{p}-\slashed{k})(\slashed{k}-k^0\gamma^0)u(p)=(2k\cdot p-k^2-2p^0k^0+\slashed{k}k^0\gamma^0)u(p) \nonumber \\ 
=(2k\cdot p-k^2-2k^0p^0+k_0^2+k^0k^z)u(p)=(-2\vec{k}\cdot \vec{p}+\vec{k}^2+k^0k^z)u(p) \ .
\end{align}
Introducing the parameters $\alpha_1=\rho(1-x)t$ for $-k^2-i0$, $\alpha_2=\rho x t$ for $(p-k)^2$, $\alpha_3=\rho(1-t)$ for $|\vec{k}|^2$, after Wick rotation one has the following representation
\begin{align}
&h_n^C=\frac{1}{\Gamma(1+n\epsilon)}\int_{0}^1 \bar x^{n\epsilon}dx \int_{0}^1 t^{1+n\epsilon}dt \int_{0}^{\infty}\rho^{n\epsilon+2}d\rho \nonumber \\ 
& \times \int \frac{d^Dk}{(2\pi)^D}e^{-\rho t \hat k_E^2-\rho \vec{k}_E^2-\rho t(1-t)x^2p_0^2}(-2xt p_0^2+\vec{k}_E^2+x^2t^2p_0^2+x^2t p_0^2) \nonumber \\ 
&=\frac{1}{\Gamma(1+n\epsilon)(4\pi)^{\frac{D}{2}}}\int_{0}^1 \bar x^{n\epsilon}dx \int_{0}^1 t^{1+n\epsilon}dt \int_{0}^{\infty}\rho^{n\epsilon+2}d\rho \left(\frac{1}{\rho t}\right)^{\frac{1}{2}}\rho^{-\frac{D-1}{2}} \nonumber \\ 
& \times e^{-p_0^2\rho t(1-t)x^2}\bigg(p_0^2t(x^2-2x)+\frac{D-1}{2\rho}+x^2t^2p_0^2\bigg) \ .
\end{align}
Now, in terms of $s=(n+1)\epsilon$, the result reads
\begin{align}
h_n^C=\frac{1}{(4\pi)^{\frac{D}{2}}}\frac{2 (\epsilon-1) s \Gamma \left(\frac{3}{2}-\epsilon\right) \left(p_0^2\right)^{-s} \Gamma (-2 s) \Gamma (-s) \Gamma (s) }{\Gamma \left(-\epsilon-s+\frac{3}{2}\right) \Gamma (-\epsilon-s+2)} \ .
\end{align}
One can see that the above simply reproduces the hard kernel obtained in the recent paper~\cite{Zhao:2023ptv} for $s=\epsilon$.
Moreover, using methods in previous sections, one can see from the factor $\Gamma(-2s)$ that the Borel transform for the renormalized hard kernel has a linear renormalon at $u=\frac{1}{2}$ as well. This renormalon is expected to be canceled by the UV renormalon for linear power corrections $O\left(\frac{1}{b_{\perp}p^z}\right)$ in TMD expansion. 
%This observation has two consequences. First, even the factorization is correct, their may still be linear divergence issues hidden somewhere, as the linear renormalon is usually associated with linear divergence. Second, even there are no linear divergence issue, the convergent speed of the ``Coulomb-gauge'' method for double-logarithmic quantities will likely to be slow. 

Here, we comment on the strength of the renormalon.  Given the above equations, the Borel transform can be extracted and reads
\begin{align}
B(u)=R(u)+\frac{C_F}{2}\frac{G_0(u)-G_{0}'(0)u-G_{0}(0)}{u^2} \ , \\ 
G_0(u)=-\frac{e^{\frac{5 u}{3}} u}{2(u-1) (2 u-1)\sin \pi u }\left(\frac{\mu^2}{4p_0^2}\right)^u \ .
\end{align}
The residue at $u=\frac{1}{2}$ reads
\begin{align}
B\bigg (u\rightarrow \frac{1}{2} \bigg)=\frac{\mu}{p_0}\frac{C_Fe^{\frac{5}{6}}}{4(u-\frac{1}{2})} \ .
\end{align}
Compared with the case for heavy-light Sudakov hard kernel (Eq.~(\ref{eq:Borel})),  the linear renormalon moves to the real part, but in the absence of a factor of $\pi$ and enhanced by a factor of $\frac{\pi}{2}$. Currently, we do not have a physical understanding of this linear renormalon, but the different transcendentality implies that the physical origin for the linear renormalons in Coulomb and axial gauges might be different.

\section{Discussion and conclusion}\label{sec:conclu}
Before ending the paper, we would like to make two comments. First,  the phase angle in the axial gauge can be {\it heuristically} explained by collinear quark exposed in an `` imaginary potential" $gA^z(t,\vec{x}_\perp)\sim i V(\sqrt{t^2-\vec{x}^2_\perp}) \sim \frac{ig^2}{\sqrt{t^2-\vec{x}_\perp^2}}$ created by the space-like gauge link, which is nothing but the analytical continuation of the standard linear potential into the time-like region $t>\vec{x}_\perp$. Since the linear potential is known to suffer from the mass renormalon, $\delta V \sim \delta m$, one expects that correction to the ``quark field renormalization'' in the external field $gA^z \sim i\delta m$ takes the form 
\begin{align}
i\frac{\slashed{p}\delta m \gamma^z}{0-2p^z}\frac{1}{2p^z}u(p)=-i\frac{\delta m}{2p^z} u(p)\ ,
\end{align}
which corresponds to the old-fashioned diagram where an anti-quark is created at $t=0$ and annihilates with the incoming quark at a large time $t>|\vec{x}_\perp|>0$.  Second, we should mention that the observed slow convergence for the imaginary part in lattice extraction of Collins Soper kernel~\cite{LatticePartonLPC:2023pdv,Avkhadiev:2023poz} can be explained by this renormalon. We hope that by using some form of Borel-resummed hard kernel taking into account the renormalon asymptotics, numerical uncertainties related to imaginary parts can be reduced and the convergence speed of perturbative matching can be improved. In our preliminary numerical test, reduction in the imaginary part has indeed been observed, see Appendix~\ref{sec:ImCSK}. 
%The same renormalon will also affect numerical uncertainty of threshold-resummation improved NNLO lattice PDF. 

To summarize, we found that the common hard kernel for double-logarithmic factorization formulas in the context of lattice parton distributions, the heavy-light Sudakov hard kernel or axial gauge quark field renormalization, has linear renormalon ambiguity in its imaginary part. In the bubble chain or large $\beta_0$ approximation, the strength of the renormalon ambiguity coincides with that for heavy-quark pole mass or heavy Wilson-line linear divergence. For both the threshold expansion of quasi-PDF matching kernel or the TMD expansion of quasi-LFWF amplitude, the linear renormalon anticipates the presence of linear power correction and correlates with UV renormalon at NLP.  The quark wave function renormalization in the Coulomb gauge also suffers linear renormalon.

\acknowledgments
We thank Prof. Xiangdong Ji for discussion. Y. Liu thanks Prof. Vladimir M. Braun for helpful discussion.  Calculations in the work are cross-checked between two authors.  Y. L. is supported by the Priority Research Area SciMat and DigiWorlds under the program Excellence Initiative - Research University at the Jagiellonian University in Krak\'{o}w. Y.S. is partially supported by the U.S.~Department of Energy, Office of Science, Office of Nuclear Physics, contract no.~DE-AC02-06CH11357.

\appendix
\section{Second Stirling numbers}\label{sec:stirling}
In this appendix, we collect useful properties and formulas related to the second Stirling numbers. As a reminder, for $n,j \in {\bf \rm N}^+$, the second Stirling number $S(j,n)$ is defined as the total number of partitions of the set $[j]=\{1,2,...j\}$ into $n$ parts. Or equivalently, $n!S(j,n)$ is the total numbers of {\it surjections} from $[j]$ to $[n]$. Clearly, $n!S(j,n)=0$ for $1 \le j<n$ (because one can not make surjection from a smaller set to a larger one) and $n!S(n,n)=n!$ (because surjections from $[n]$ to $[n]$ are permutations). On the other hand, for  $ 1\le i \le n$, in terms of the sets $S_i=\{f \in [n]^j, i\notin {\rm Im}(f)\}$, one actually has
\begin{align}
n!S(j,n)&=\# \bigg([n]^j\bigg /\bigcup_{i=1}^n S_i\bigg)=n^j+\sum_{k=1}^{n}(-1)^k\sum_{1\le i_1<i_2..<i_k\le n}\# \bigcap_{l=1}^k S_{i_l} \nonumber \\ 
&=\sum_{k=0}^n (-1)^k\binom{n}{k}(n-k)^j \ .
\end{align}
In the first line we used the standard inclusion-exclusion principle and in the second line we used the fact that the total number of $1\le i_1<i_2<..i_k\le n$ is $\binom{n}{k}$, and for each of them the size of the set $\bigcap_{l=1}^k S_{i_l}$ is $(n-k)^j$. As a result, one has
\begin{align}
    &\sum_{k=0}^n (-1)^k\binom{n}{k}(n-k)^j=0, \  0 \le j\le n-1 \ , \\ 
    &\sum_{k=0}^n (-1)^k\binom{n}{k}(n-k)^n=n! \ .
\end{align}
Given the above, Eq.~(\ref{eq:S(n,j)}) and Eq.~(\ref{eq:S(n,n)}) follows immediately after expanding $(n-k+x)^j$ and $(n-k+x)^n$. On the other hand, using $\binom{n}{k}=\frac{k+1}{n+1}\binom{n+1}{k+1}$, 
one has
\begin{align}
\sum_{k=0}^{n}(-1)^k \binom{n}{k}(n-k+1)^{-1}=\frac{1}{n+1}\sum_{k=0}^{n}(-1)^{n-k}\binom{n+1}{k+1}=\frac{(-1)^n}{n+1} \ ,
\end{align}
while
\begin{align}
\sum_{k=0}^{n}(-1)^k \binom{n}{k}(n-k+1)^{-j}=n!S(-j+1,n+1) \ , \ j\ge 2 \ .
\end{align}
Here $S(-j+1,n+1)$ are called ``{\it negative-positive}'' second Stirling numbers in the literature. They can be represented as
\begin{align}
S(-j+1,n+1)=\frac{(-1)^{n+j-1}}{(j-1)!}\frac{d^{j-1}}{dx^{j-1}}\bigg[\frac{\Gamma(x+1)}{\Gamma(x+n+2)}\bigg]_{x=0} \ .
\end{align}
This implies that 
\begin{align}
&\sum_{k=0}^{n}(-1)^k \binom{n}{k}(n-k+1)^{-2}=(-1)^n\frac{H_{n+1}}{n+1} \ , \\
&\sum_{k=0}^{n}(-1)^k \binom{n}{k}(n-k+1)^{-3}=(-1)^n\frac{H_{n+1}^2+H^{(2)}_{n+1}}{2(n+1)} \ , \\ 
&\sum_{k=0}^{n}(-1)^k \binom{n}{k}(n-k+1)^{-4}=(-1)^n\frac{2H_{n+1}^{(3)}+3H_{n+1}H_{n+1}^{(2)}+H_{n+1}^3}{6(n+1)} \ .
\end{align}
Here $H^{(i)}_n=\sum_{k=1}^n\frac{1}{k^i}$ are the higher harmonic numbers. 
The last two formulas are required in the presence of $\frac{1}{\epsilon^3}$ and $\frac{1}{\epsilon^4}$ poles. 
\section{Absence of linear renormalon in the light-light Sudakov hard kernel}
In this appendix, for the convenience of the readers, we show that the standard light-light Sudakov hard kernel for $j^{\mu}=\bar \psi \gamma^{\mu} \psi$ is free from linear renormalon ambiguity.

We calculate the bubble chain diagram for the vertex correction of the light light Sudakov factor. The diagram dressed up with $n$ bubbles is
\begin{equation}
    \bar{u}(p') C_n\gamma^{\rho} u(p) = \alpha 4 \pi C_F \mu^{2 \epsilon} \left(\frac{e^{\gamma_E}}{4\pi}\right)^{\epsilon} \bar{u}(p') \int \frac{d^{d}k}{(2\pi)^d} D^{(n)}_{\mu \nu}(k) \gamma^{\mu} \frac{1}{\slashed{p'}-\slashed{k}} \gamma^{\rho} \frac{1}{\slashed{p}-\slashed{k}} \gamma^{\nu} u(p),
\end{equation}
where the calculation is performed under $d=4-2\epsilon$ dimension. $u(p)$ and $\bar{u}(p')$ are the Dirac spinors with $p^2=p'^2=0$. We define $q=p'-p$ and $Q^2=-q^2$. $D^{(n)}_{\mu \nu}(k)$ is the gluon propagator with $n$ bubbles
\begin{equation}
D^{(n)}_{\mu \nu}(k) = \frac{-i \left(g_{\mu \nu}-\frac{k_{\mu}k_{\nu}}{k^2}\right)}{k^2} \left(3 \beta_0 \alpha \left(\frac{\mu^2}{-k^2}\right)^{\epsilon} e^{\epsilon \gamma_E} \frac{\Gamma(2-\epsilon)^2 \Gamma(\epsilon)}{\Gamma(4-2\epsilon)}\right)^n,
\end{equation}
where $\beta_0 = \frac{11 C_A}{6 \pi} - \frac{n_f}{3\pi}$. Using Feynman and Schwinger parameters, we obtain
\begin{equation}
    C_n= 
    \frac{C_F}{2 \beta_0} \left(\frac{\alpha \beta_0}{2}\right)^{n+1} \frac{1}{(n+1)^2 \epsilon^{n+2}} G(\epsilon,(n+1)\epsilon) \ .
\end{equation}
where the UV and IR poles are not distinguished in the above expression. $G(\epsilon,s)$ is analytic at $\epsilon,s=0$,  
\begin{equation}\label{eq:Ges}
G(\epsilon,s)=\left(\frac{\mu ^2}{Q^2}\right)^s e^{\gamma_E  s} f(\epsilon)^{\frac{s}{\epsilon }-1} \frac{2 s \left(s^2 \epsilon +s \epsilon ^2-2 s \epsilon -\epsilon ^2+3 \epsilon -2\right) \csc (\pi  s) \Gamma (1-s)}{\Gamma (-s-\epsilon +3)} \ .
\end{equation}
One can do Taylor expansion at $s=0$
\begin{equation}\label{eq:Gexpand}
    G(\epsilon,s)=\sum_{j=0}^{+\infty} G_{j}(\epsilon) s^j,
\end{equation}
where $G_{j}(\epsilon)$ is analytic at $\epsilon=0$. Now, using the standard method, the partially renormalized result is
\begin{align}
    &R_{\rm part}C_n =\frac{C_F}{2\beta_0} \left( \frac{\alpha\beta_0}{2}\right)^{n+1} \left( \frac{G_0(\epsilon)}{\epsilon^{n+2}} \frac{(-1)^n H_{n+1}}{n+1} + \frac{G_1(\epsilon)}{\epsilon^{n+1}} \frac{(-1)^n}{n+1} + G_{n+2}(\epsilon) \, n! \right) + O(\epsilon) \ ,
\end{align}
and the fully renormalized result in $\overline{\rm MS}$ scheme is
\begin{align}\label{eq:LLfullR}
    &RC_n =\frac{C_F}{2\beta_0} \left( \frac{\alpha\beta_0}{2}\right)^{n+1} \left( \frac{ \frac{d^{n+2}G_0(\epsilon)}{d \, \epsilon^{n+2}} |_{\epsilon \rightarrow 0}}{(n+2)!} \frac{(-1)^n H_{n+1}}{n+1} + \frac{ \frac{d^{n+1}G_1(\epsilon)}{d \, \epsilon^{n+1}} |_{\epsilon \rightarrow 0}}{(n+1)!} \frac{(-1)^n}{n+1} + G_{n+2}(\epsilon=0) \, n! \right) \ .
\end{align}
In Eq.~(\ref{eq:LLfullR}), $\frac{ \frac{d^{n+2}G_0(\epsilon)}{d \, \epsilon^{n+2}} |_{\epsilon \rightarrow 0}}{(n+2)!}$ and $\frac{ \frac{d^{n+1}G_1(\epsilon)}{d \, \epsilon^{n+1}} |_{\epsilon \rightarrow 0}}{(n+1)!}$ don't contribute to $n!$ growths since $G_j(\epsilon )$ is analytic at $\epsilon=0$. Only the last term $G_{n+2}(0) \, n!$ contains a factorial growth. Thus the asymptotic form in the large $n$ limit is
\begin{align}
    RC_{n,{\rm asym}} &= \frac{C_F}{2\beta_0} \left( \frac{\alpha\beta_0}{2}\right)^{n+1} G_{n+2}(0) \, n! = p^{\rm asym}_n \alpha^{n+1},
\end{align}
where $p^{\rm asym}_n = \frac{C_F}{2\beta_0} \left( \frac{\beta_0}{2}\right)^{n+1} G_{n+2}(0) n!$. The Borel transformation is
\begin{align}
   &B(u) = \frac{C_F}{4u^2} \left( -\frac{4ue^{\frac{5 u}{3}} \csc \left(\pi u\right) \Gamma \left(1-u\right) \left(\frac{\mu
   ^2}{Q^2}\right)^{u}}{\Gamma \left(3-u\right)} + \frac{2}{\pi } + \frac{u\left(3 \ln \left(\frac{\mu ^2}{Q^2}\right)+19\right)}{6 \pi }  \right) \ .
\end{align}
The above expression is analytical at $u=\frac{1}{2}$ thus the light light Sudakov form factor doesn't suffer from linear renormalon.

The light light Sudakov hard kernel in the large $\beta_0$ limit is
\begin{align}
    C= 1 + \sum_{n=0}RC_n \ .
\end{align}
Following the same strategy as the heavy light Sudakov form factor, one can calculate the anomalous dimensions in the large $\beta_0$ limit as follows
\begin{align}
    &\frac{d C}{d \ln \mu} = -\frac{C_F}{\beta_0} \int_{0}^{1} d t \frac{G_{0}(u) - G_{0}(u t)}{1-t} + \frac{C_F \alpha}{2} \tilde{G}(u) + \frac{C_F \alpha}{2} G_{0}(u) \ln \frac{\mu^2}{Q^2} \nonumber\\
    &= \left(-\frac{C_F}{\beta_0} \int_{0}^{1} d t \frac{G_{0}(u) - G_{0}(u t)}{1-t} + \frac{C_F \alpha}{2} \tilde{G}(u) + \frac{C_F \alpha}{2} G_{0}(u) \ln \frac{\mu^2}{Q^2}\right) C + O\left(\frac{1}{\beta_0^2}\right) \ ,
\end{align}
where $u = - \frac{\alpha \beta_0}{2}$. $\tilde G(u) \equiv G_{1}(u) - G_{0}(u) \ln \frac{\mu^2}{Q^2}$ and $\tilde G(u)$ is independent of $\ln \frac{\mu^2}{Q^2}$. Thus the cusp anomalous dimension is
\begin{align}
    \Gamma_{\rm cusp} = - \frac{C_F \alpha}{2} G_{0}(u) = \frac{2^{\alpha  \beta _0+3} C_F \sin \left(\frac{1}{2} \pi  \alpha  \beta _0\right) \Gamma \left(\frac{1}{2}
   \left(\alpha  \beta _0+5\right)\right)}{3 \pi ^{5/2} \beta _0 \Gamma \left(\frac{\alpha  \beta _0}{2}+2\right)} \ ,
\end{align}
which agrees with Eq.~(\ref{eq:cusp}). The small $\alpha$ expansion (one also replaces $\beta_0 \rightarrow -\frac{n_f}{3\pi}$) reproduces the cusp anomalous dimension in the large $n_f$ limit up to NNNLO
\begin{align}
\Gamma_{\rm cusp} = \frac{\alpha  C_F}{\pi } -\frac{5 \alpha ^2 C_F n_f}{18 \pi ^2} - \frac{\alpha ^3 C_F n_f^2}{108 \pi ^3}  \ ,
\end{align}
which is consistent with that from~\cite{Becher:2006mr} in large $n_f$ limit. The single log anomalous dimension is
\begin{align}
    \gamma_V = -\frac{C_F}{\beta_0} \int_{0}^{1} d t \frac{G_{0}(u) - G_{0}(u t)}{1-t} + \frac{C_F \alpha}{2} \tilde{G}(u) \ ,
\end{align}
whose small $\alpha$ expansion reproduces the single log anomalous dimension in the large $n_f$ limit up to NNNLO
\begin{align}
\gamma_V = -\frac{3 \alpha  C_F}{2 \pi } + \frac{\left(65+9 \pi ^2\right) \alpha ^2 C_F n_f}{216 \pi ^2}-\frac{\alpha ^3 \left(216 \zeta (3)-2417+270 \pi^2\right) C_F n_f^2}{23328 \pi ^3} \ ,
\end{align}
which is consistent with that from~\cite{Becher:2006mr} in large $n_f$ limit. Setting $\beta_0 \rightarrow -\frac{n_f}{3\pi}$ and $\mu=Q$ in Eq.~(\ref{eq:LLfullR}), one can also obtain the constants term in the large $n_f$ limit,
\begin{align}
C = 1 +\frac{\alpha  C_F}{4 \pi } \left(\frac{\pi ^2}{6}-8\right) + \left(\frac{\alpha}{4\pi}\right)^2 C_F n_f T_F \left(\frac{4 \zeta (3)}{9}+\frac{4085}{162}+\frac{23 \pi ^2}{27}\right) \ ,
\end{align}
which is consistent with that from~\cite{Becher:2006mr} in large $n_f$ limit.

\section{``Imaginary part" of Collins-Soper kernel}\label{sec:ImCSK}
The Collins-Soper (CS) kernel describes the rapidity dependence of the TMDPDF or TMDWF~\cite{Collins:1981uk,Collins:1981va}. It was proposed that the CS kernel can be extracted on lattice through quasi-TMDPDF~\cite{Ebert:2018gzl} or quasi-TMDWF~\cite{Ji:2021znw} with perturbative matching. The CS kernel is purely real by definition. 

However, a non-vanishing imaginary part was observed during the extraction of the CS kernel through quasi-TMDWF~\cite{LPC:2022ibr,LatticePartonLPC:2023pdv,Avkhadiev:2023poz}. According to these papers, the ratio of the quasi-TMDWFs is nearly real and the matching correction is a dominant source of the imaginary part of the CS kernel.

\begin{figure}[htbp]
    \centering
    \includegraphics[height=6.18cm]{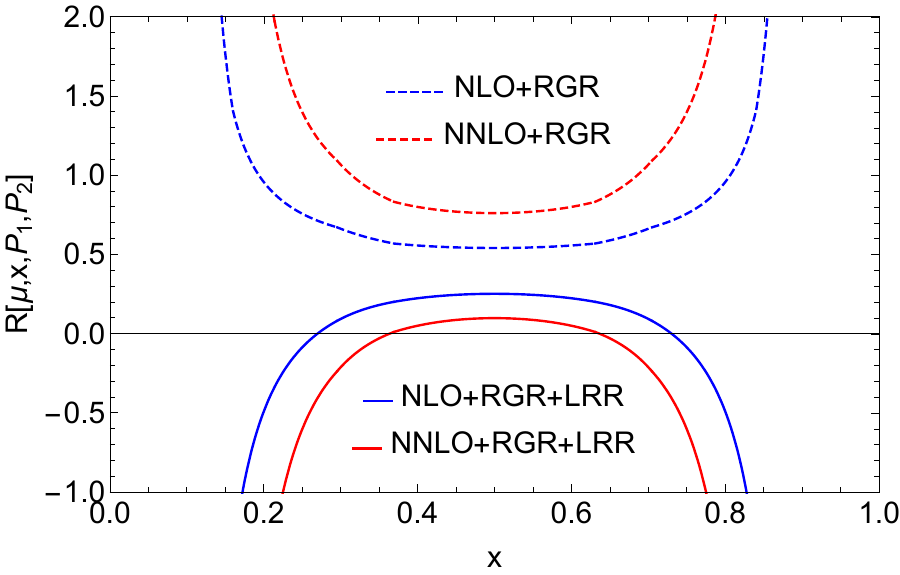}
    \caption{The imaginary part of the matching correction during extracting the CS kernel through quasi-LFWF amplitude. The dashed curves are the results only with RGE resummation. The solid curves are the results with the RGE and leading renormalon resummation (LRR). We choose the typical parameters under current lattice techniques, $\mu=2$ GeV, $P_1=1.72$ GeV, $P_2=2.15$ GeV and $n_f = 3$. }
    \label{fig:Rfactor}
\end{figure}

To study the imaginary part of the matching correction, we define a $R$ factor following the similar strategy in Appendix C in Ref.~\cite{Avkhadiev:2023poz},
\begin{align}
    &R[\mu, x, P_1, P_2] = \frac{1}{2\ln P_1/P_2} \left[ A\bigg(\ln\frac{4 x^2 P_1^2}{\mu^2},\alpha(\mu)\bigg) + A\bigg(\ln\frac{4 (1-x)^2 P_1^2}{\mu^2},\alpha(\mu)\bigg) \right.\nonumber\\
    &\left.- A\bigg(\ln\frac{4 x^2 P_2^2}{\mu^2},\alpha(\mu)\bigg) - A\bigg(\ln\frac{4 (1-x)^2 P_2^2}{\mu^2},\alpha(\mu)\bigg) \right]
\end{align}
where $A$ is the phase angle defined in Eq.~(\ref{eq:defJf}). We perform the renormalization group resummation (RGR) for the phase angle $A$ based on Eq.~(\ref{eq:phaseRG}). The $R$ factors with the RG resummed phase angles are shown as the dashed curves in Fig.~\ref{fig:Rfactor}. The dashed curves deviate from zero, which causes the non-vanishing imaginary part for the CS kernel. Similar observations have been found in Ref.~\cite{Avkhadiev:2023poz}.

We make use of the asymptotic form Eq.~(\ref{eq:asympAb}) to perform the leading renormalon resummation (LRR)~\cite{Zhang:2023bxs} for the phase angle $A$. The $R$ factors with the RGR+LRR phase angles are shown as the solid curves in Fig.~\ref{fig:Rfactor}. After the leading renormalon resummation, the imaginary part of the matching correction is suppressed during the moderate $x$ range. Thus the imaginary part of the CS kernel is expected to be suppressed with leading renormalon resummed matching coefficient. 

In the above LRR approach, we regulate the renormalon divergence in Borel plane with the principle value prescription~\cite{Cvetic:2002qf}. One can choose different contours in Borel plane, corresponding to different results with the renormalon ambiguity at $O\left( \frac{\Lambda_{\rm QCD}}{x (1-x) P_1}-\frac{\Lambda_{\rm QCD}}{x (1-x) P_2} \right)$. According to Eqs.~(\ref{eq:NLPresults})(\ref{eq:NLPresidue}), this renormalon ambiguity is mixed with the NLP TMDWF. Thus, to achieve the leading power accuracy for the extraction of CS kernel through TMDWF, one needs to consider NLP TMDWF. One approach is to parametrize the NLP TMDWF and perform some phenomenological fits with lattice data, which requires further investigation.

\section{The $U_{C}^{\dagger}(\vec{x})$ $U_{C}(0)$ correlator in Coulomb gauge}
The Coulomb gauge quasi-PDF was proposed in Ref.~\cite{Gao:2023lny}, which is expected to be free from the linear divergence and the associated renormalon. The numerical tests in Ref.~\cite{Gao:2023lny} indicate that there is no linear divergence in the lattice matrix elements. To check the linear renormalon, we calculate the bubble chain diagram for the $U_{C}^{\dagger}(\vec{x})$ $U_{C}(0)$ correlator, where $U_{C}(\vec{x})$ is defined in Eq.~(\ref{eq:CGDS}).

The $U_{C}^{\dagger}(\vec{x})$ $U_{C}(0)$ correlator is similar to the self-energy diagram for quasi-PDF in standard gauge and will also be encountered in the TMD objects playing the role of the ``gauge-link self-interactions''.  The bubble chain diagram with $n$ bubbles is
\begin{align}
g^2C_F\alpha^n\bigg(\frac{\beta_0}{2}\bigg)^n\mu_0^{2\epsilon}\mu^{2n\epsilon}e^{n\epsilon \gamma_E}\frac{1}{\epsilon^n}f(\epsilon)^n f_n(\vec{x}),
\end{align}
where $f(\epsilon)$ is defined in Eq.~(\ref{eq:feps}). The relevant Feynman integral $f_n(\vec{x})$ is 
\begin{align}
&f_n(\vec{x})=-i\int \frac{d^Dk}{(2\pi)^D} \frac{e^{i\vec{k}\cdot \vec{x}}}{|\vec{k}|^2}\frac{1}{(-k^2-i0)^{1+n\epsilon}}\nonumber \\ 
&=\frac{1}{\Gamma(1+n\epsilon)}\int \frac{d^Dk}{(2\pi)^D} e^{i\vec{k}\cdot \vec{x}}\int_{0}^{\infty} d\alpha_1d\alpha_2 \alpha_1^{n\epsilon} e^{-\alpha_1 k^2-\alpha_2\vec{k}^2} \nonumber \\
&=\frac{1}{\Gamma(1+n\epsilon)}\int \frac{d^Dk}{(2\pi)^D} \int_{0}^{\infty} \rho^{1+n\epsilon}d\rho\int_{0}^1x^{n\epsilon}dxe^{-\rho x k_0^2-\rho \vec{k}^2+i\vec{k}\cdot \vec{x}} \nonumber \\ 
&=\frac{1}{(4\pi)^{\frac{D}{2}}\Gamma(1+n\epsilon)}\int_{0}^{\infty} \rho^{(n+1)\epsilon-1}d\rho\int_{0}^1x^{n\epsilon-\frac{1}{2}}dxe^{-\frac{\vec{x}^2}{4\rho}} \nonumber \\ 
&=\frac{1}{(4\pi)^{\frac{D}{2}}\Gamma(s-\epsilon+1)}\frac{2^{1-2 s} \left(\vec{x}^2\right)^{s} \Gamma (-s)}{-2 \epsilon+2 s+1} \ .
\end{align}
As before, $s=(n+1)\epsilon$. Based on the standard procedure, it is free from linear renormalon.

\bibliographystyle{apsrev4-1} 
\bibliography{bibliography}

\end{document}